\documentclass{emulateapj}
\usepackage{graphicx}
\usepackage{longtable}
\usepackage{pdflscape}

\def\TMB {$T_{\rm mb}$}

\def\TASTAR {$T_{\rm A}^{*}$}

\def\H0 {$H_{\rm o}$}

\def\solum {\hbox{L$_{\odot}$}}

\def\kms {\hbox{${\rm km\,s}^{-1}$}}

\def\Kkms {\hbox{${\rm K\,km\,s}^{-1}$}}
\def\percc {$\hbox{{\rm cm}}^{-3}$}    

\def \apj {{ ApJ}}
\def \apjl {{ ApJ}}
\def \apjs {{ ApJS}}
\def \pasj {{ PASJ}}
\def \araa {{ ARA\&A}}

\def \aap {{ A\&A}}
\def \aj {{ AJ}}
\def \mnras {{ MNRAS}}

\shorttitle{An extragalactic CO(3--2) survey}

\shortauthors{Mao, Schulz, Henkel et al.}

\slugcomment{Accepted for publication in The Astrophysical Journal}

\begin{document}

\title
{An Extragalactic $^{12}$CO~$J$~=~3--2 survey with the
Heinrich-Hertz-Telescope}

\author
{Rui-Qing~Mao\altaffilmark{1,2},
 Andreas~Schulz\altaffilmark{3,4},
 Christian Henkel\altaffilmark{2},
 Rainer~Mauersberger\altaffilmark{5}, Dirk~Muders\altaffilmark{2}
 and Dihn-V-Trung\altaffilmark{6,7}}

\altaffiltext{1}{Purple Mountain Observatory, Chinese Academy of
 Sciences, 210008 Nanjing, PR China;  {\tt rqmao@pmo.ac.cn}}
\altaffiltext{2}{Max-Planck-Institut f{\"u}r Radioastronomie,
 Auf dem H{\"u}gel 69, D-53121 Bonn, Germany}
\altaffiltext{3}{ Argelander-Institut f\"ur Astronomie,
 Universit\"at Bonn, Auf dem H\"ugel 71, D-53121 Bonn, Germany}
\altaffiltext{4} {Institut f{\"u}r Physik und ihre Didaktik,
 Universit{\"a}t zu K{\"o}ln, Gronewaldstr. 22, D-50931 K{\"o}ln,
 Germany}
\altaffiltext{5}{Joint ALMA Observatory, Av. Alonso de C{\'o}rdova 3107 ,
 Vitacura, Santiago, Chile}
\altaffiltext{6}{Institute of
 Astronomy and Astrophysics, Academia Sinica, Taipei, Taiwan}
\altaffiltext{7}{Center for Quantum Electronics, Institute of
 Physics, Vietnamese Academy of Science and Technology, 10
 DaoTan, BaDinh, Hanoi, Vietnam}

\begin{abstract}
We present results of a $^{12}$CO $J$ = 3--2 survey of 125
nearby galaxies obtained with the 10-m Heinrich-Hertz-Telescope,
with the aim to characterize the properties of warm and dense
molecular gas in a large variety of environments. With an angular
resolution of 22$''$, $^{12}$CO 3--2 emission was detected in 114
targets. Based on 61 galaxies observed with equal beam sizes the
$^{12}$CO 3--2/1--0 integrated line intensity ratio $R_{\rm 31}$
is found to vary from 0.2 to 1.9, with an average value of
0.81. No correlations are found for $R_{\rm 31}$ to
Hubble type and far infrared luminosity. Possible indications for
a correlation with inclination angle and the 60$\mu$m/100$\mu$m
color temperature of the dust are not significant. Higher $R_{\rm 31}$
ratios than in ``normal'' galaxies, hinting at enhanced molecular
excitation, may be found in galaxies hosting active galactic nuclei.
Even higher average values are determined for galaxies with bars or
starbursts, the latter being identified by the ratio of infrared
luminosity versus isophotal area,
log\,[($L_{\rm FIR}/L_{\odot}$)/($D_{25}^2$/kpc$^2$)]
$>$ 7.25. (U)LIRGs are found to have the highest averaged $R_{\rm 31}$
value. This may be a consequence of particularly vigorous star formation
activity, triggered by galaxy interaction and merger events. The nuclear
CO luminosities are  slightly sublinearly correlated with the global
FIR luminosity in both the $^{12}$CO $J$ = 3--2 and the 1--0 lines. The 
slope of the log-log plots rises with compactness of the respective
galaxy subsample, indicating a higher average density and a larger
fraction of thermalized gas in distant luminous galaxies. While
linear or sublinear correlations for the $^{12}$CO $J$ = 3--2 line
can be explained, if the bulk of the observed $J$ = 3--2 emission
originates from molecular gas with densities below the critical
one, the case of the $^{12}$CO $J$ = 1--0 line with its small
critical density  remains a puzzle.

\end{abstract}


\section{Introduction}
\label{section:intro}

Low lying rotational transitions of CO are widely used as tracers
of molecular hydrogen and are essential to determine dynamical
properties and total molecular masses of galaxies. The widespread
use of $^{12}$CO $J$ = 1--0 and 2--1 (hereafter CO(1--0) and
CO(2--1)) spectroscopy
\citep[e.g.][]{bra93,you95,chi96,elf96,alb04,alb07} is, however,
not sufficiently complemented by systematic surveys in higher
rotational CO transitions to constrain the excitation conditions
of the dense interstellar medium (ISM). While the $J$ = 1 and 2
states of CO are only 5.5 and 17\,K above the ground level, the
$J$ = 3 state is at 33\,K and traces a component of higher
excitation. The ``critical densities'', at which collisional
de-excitation matches spontaneous decay in the optically thin
limit, is $\sim$ 10$^{5}$\,\percc\ for $^{12}$CO $J$ = 3--2
(hereafter CO(3--2)) in contrast to 10$^{3.5}$ and
10$^{4.3}$\,\percc\ for the two lower rotational CO transitions.
Therefore, the CO(3--2) line is a particularly useful tracer of
the molecular gas properties in the central regions of galaxies,
where the molecular gas is generally believed to be warmer and
denser than in typical galactic disk clouds \citep[e.g.,][]{gue81,mau93}. 
The CO(3--2) to (1--0) line
intensity ratio is better suited to constrain the gas temperature
and density than the ratio of CO(2--1) to (1--0).

In the local universe, most of the evidence for a higher excited
gas phase comes from species other than CO. In many cases,
however, such rare molecular species are difficult to detect,
particularly in higher excited transitions. To investigate
properties (e.g. spatial density, column density, kinetic
temperature) of the bulk of the gas for a large sample of
galaxies, observations of strong lines are needed. Thus, CO
transitions of higher excitation have to be observed. This has
been proved to be very successful in the first extragalactic
CO(3--2) survey encompassing a significant (29) number of galaxies
\citep{mau99}, in which CO(3--2) was detected in all of the
targets studied.

Encouraged by this result \citep[see also][]{dev94}, we
have used the Heinrich Hertz Telescope (HHT) on Mt. Graham
\citep{baa96} to observe the CO(3--2) line in an extended sample
of galaxies. After we started this extended project,
additional extragalactic CO(3--2) surveys have been carried out,
aiming at different types of galaxies. These include Virgo cluster
galaxies \citep{haf03}, infrared luminous galaxies
\citep{yao03,nar05}, early type galaxies \citep{vil03}, compact
and dwarf galaxies \citep{mei01,isr05}, double barred galaxies
\citep{pet03,pet04}, nearby galaxies of various types
\citep{bay06}, and most recently some nearby spiral and elliptical
galaxies with two  newly mounted submillimeter telescopes in
Chile, APEX and ASTE \citep{nak07,kom07,gal08}. Some extended
CO(3--2) maps were also reported toward a handful of nearby
galaxies \citep[e.g.,][]{dum01,mur09,war10}. Interferometric
CO(3--2) maps of individual galaxies are also feasible thanks
to the Submillimeter Array \citep[hereafter SMA; see,
e.g.,][]{wil08,wil09}. However, the detailed study of warm gas in
nearby galaxies is still in its infancy. This is regrettable,
in particular because of its crucial role in highly redshifted
targets, where not CO(1--0) but higher excited CO lines are
commonly observed \citep[e.g.,][]{sol05}.

In this paper, we present the results of our survey, which
covers the by far largest sample of galaxies measured so far in
the CO(3--2) line. The data were obtained in a ``coherent'' way,
making use of a specific telescope/receiver/backend combination.
With complementary information from CO(1--0), taken from the
literature, we thus present a data base of unprecedented
size, providing a suitable basis to check the quality of our data
(as well as those reported earlier) and allowing us to tackle
a number of astrophysical questions. The data are used for the
following main purposes:  1) to provide a large and homogeneous
data set of CO(3--2) spectra, which will form an essential basis
for future studies, either aiming at higher angular resolution
or searching for higher excited CO transitions see, e.g.,
\citep{vdw10} for a pioneering study), 2) to systematically
trace the global properties of the warm and dense molecular gas
in various galaxies, 3) to test whether there are any
correlations between the molecular gas excitation (given
by the CO(3--2)/CO(1--0) intensity ratio) and galaxy properties
such as Hubble type, nuclear activity, far infrared
(FIR:40--400~$\mu$m) luminosity, 60$\mu$m/100$\mu$ color
temperature of the dust, and inclination, 4) to evaluate the
effect of galaxy interactions on the molecular gas properties,
5) to test whether the CO(3--2) line is a better tracer of star
formation than the CO(1--0) line, and 6) to evaluate the
Schmidt-Kennicutt law in the light of the new data. We present
the sample selection in \S2, the observations in \S3, the basic
results in \S4, a systematic analysis of correlations
in \S5, and the summary in \S6.

\section{The samples}
\label{section:sample}

\subsection{Sample selection} Our sample selected for the CO(3--2)
survey consists of 125 galaxies which are part of five major
partially overlapping sub-samples. Table~\ref{tbl:basic} lists all
the sample galaxies along with some basic properties mostly drawn
from the NED\footnote{The NASA/IPAC Extragalactic Database (NED) is
operated by the Jet Propulsion Laboratory, California Institute of
Technology, under contract with the National Aeronautics and Space
Administration.} and HyperLEDA\footnote{HyperLEDA database: {\tt
http://leda.univ-lyon1.fr}} \citep{pat03}.

The first sub-sample consists of 58 nearby galaxies of various types. It
contains 22 reobserved sources that were already part of our initial survey
\citep{mau99}, and is complemented by the remaining IRAS point
sources with $S_{100\,\mu m} > 50$\,Jy and $\delta > -30^{\circ}$
\citep{hen86} as well as sources observed by \citet{bra93} in the
CO (1--0) and (2--1) transitions (with the IRAM 30-m telescope) if
integrated intensities are $\geq$~10\,K\,\kms.

The second sub-sample consists of 32 galaxies from a volume limited
sample ($V<$7000\,\kms) of all Seyfert galaxies and low-ionization
nuclear emission-line regions (LINERs) in Huchra's catalog of AGN
\citep{huc93} or in the \citet{ver91} catalog
 that are also included in the {\it Revised Shapley Ames
Catalog} (RSA) and that are accessible with the HHT (74 in total).

The third sub-sample consists of 25 early type galaxies from
\citet{hen97} with CO(1--0) and/or CO(2--1) lines detected. Observations
at radio, optical, and X-ray wavelengths have shown that early-type
galaxies contain an interstellar medium (ISM) comprising the same
components as found in spiral galaxies, but with different mass
fractions of the gas components (see also \citeauthor{hen97}
\citeyear{hen97} for a review).

The fourth sub-sample consists of clearly identified 11
interacting or merging systems, which are mainly luminous infrared
galaxies (LIRGs, 10$^{11}$~$\solum$~$\leq$~$L_{\rm
FIR}$~$<$~10$^{12}$\,L$_{\odot}$). The selection is based on their
relatively high single dish and/or interferometer CO(1--0) fluxes
\citep[see, e.g.,][and references therein]{san91,gao99,lo00}.
These galaxy systems are thought to be at different
merging/interaction phases, i.e. at the early (presumably
pre-starburst: Arp~303N/S, UGC~8335A/B, NGC~5257/8, Arp~302N/S,
Arp~293 and NGC~6670A/B), intermediate (Arp~55, Mrk~848 and
NGC~4038/9), or late stages of interaction (NGC~1614, NGC~5256),
according to the spatial separation of the respective galaxy pair
in each system. The two core positions of five early mergers (not
Arp~293) as well as NGC~4038/9 (the Antennae) were measured separately
because they are spatially resolved by our 22$''$ beam (see \S3).
Including these double core positions, the fourth sub-sample contains
a total of 17 individual sources. For the central position of
Arp~302, ``Arp~302~center'' (Table~\ref{tbl:basic}), see also
\S A.2.

The members of the fifth sub-sample are prominent OH megamaser
galaxies, including 4 ultraluminous infrared galaxies (ULIRGs,
$L_{\rm FIR}$~$\geq$ 10$^{12}$\,L$_{\odot}$: IRAS~17208-0014,
Mrk~231, Mrk~273, and Arp~220) and 2 LIRGs (III~ZW~35 and NGC~3690B),
all of them being late mergers (following the classification
outlined above), except NGC~3690B, which is ``early''.

\subsection{Sample properties as a whole}
\label{section:sample-pro}

\subsubsection{The IRAS fluxes}
\label{section:IRAS}

98 of our sample galaxies are part of the $IRAS$ Revised Bright
Galaxy Sample (RGBS) by \citet{san03}, which provides revised
IRAS fluxes. 23 of these were determined with particularly high
precision, also profiting from the HIRES imaging reconstruction
technique \citep{sur04}. With the higher spatial
resolution obtained by this technique, four of our galaxy pairs
(i.e., Arp~303N/S, NGC~5257/8, UGC~8335A/B, and Arp~302N/S) are
resolved and fluxes for each individual galaxy are available.
The mid- to far-IR emission of the NGC~4038/9 system (the Antennae)
originates predominantly from the overlap region where the disks
of two galaxies interact \citep[e.g.,][]{sch07}. Flux densities
directly obtained from IRAS catalogs, i.e. {\it the IRAS Point Source
Catalog} \citep[PSC;][]{psc}, {\it the IRAS Explanatory Supplement}
\citep{bei88}, and {\it the IRAS Faint Source Catalog}
\citep[FSC;][]{mos92} were taken for the rest of the sample.
We then applied the flux densities to calculate the FIR
luminosity ($L_{\rm FIR}$ = $L$(40--400~$\mu$m), following
\citet{mos92}, and the 60~$\mu$m/100~$\mu$m color temperature
($T_{\rm dust}$), assuming an emissivity that is proportional
to the frequency $\nu$. Two galaxies, IC~750 and NGC~4138,
were not observed by IRAS.

\subsubsection{Galaxy classifications}
\label{section:classification}

With improved observations, galaxy classifications may have to
be modified in some cases. We have used the NED classifications
from August 2008 as standard throughout the paper. While
Seyferts/LINERs can be directly recognized from NED,
starbursts are not explicitly indicated. There exists a variety
of definitions of the starburst phenomenon in the literature,
which have been reviewed by, e.g., \citet[][]{hec05} and
\citet[][]{ken05}. A starburst can be defined in terms of
its absolute star forming rate (SFR), its SFR surface density
(the SFR per unit area), or if its SFR exceeds an average value
from the past by a fixed amount. The situation is further complicated
by the choice of the respective SFR tracer like, e.g., the
ultraviolet emission, the far infrared emission, or the radio
continuum. Spectroscopic tracers like recombination lines (e.g.,
H$\alpha$) have also been frequently used.

\begin{figure*}
\centering
\includegraphics[angle=0,width=14cm]{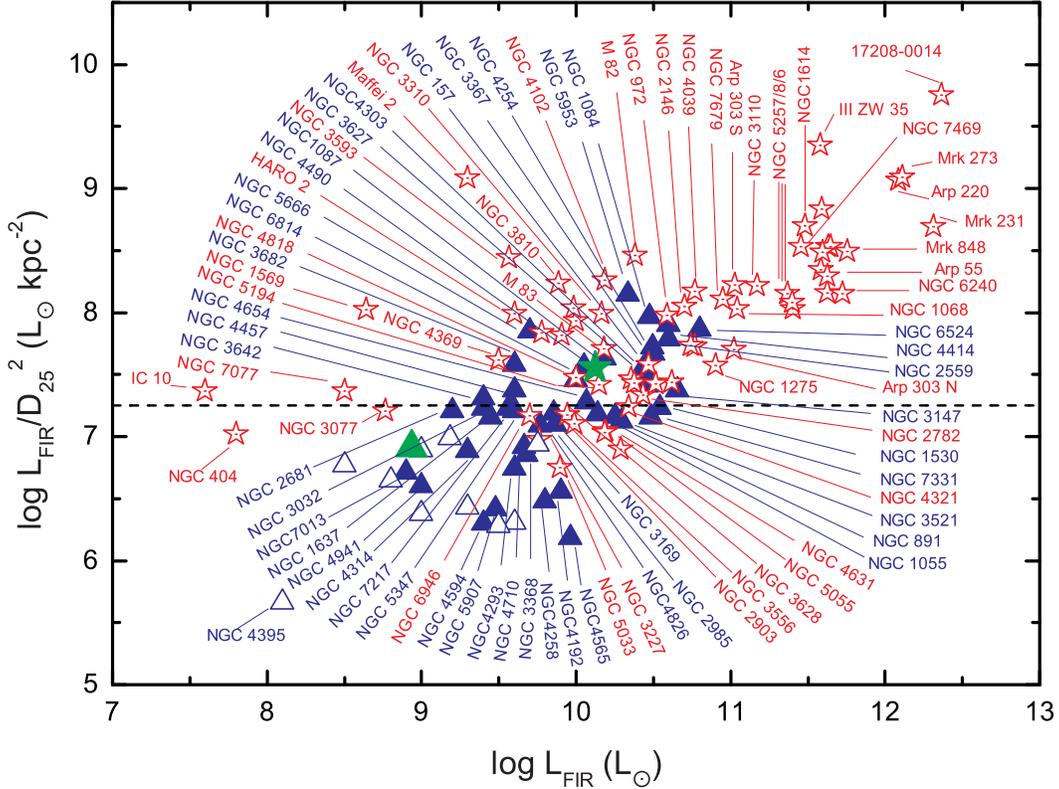}
\caption[]{Starburst definition: Red open stars denote galaxies
classified as starbursts in the literature. Blue triangles mark
galaxies not being classified as such. All our CO $J$ = 3--2
non-detections arise from the latter sample and are marked by open
triangles. Also marked are NGC\,253 (by a green star), a typical
starburst galaxy \citep[e.g.,][]{bru09}, and IC\,342 (by a green
triangle), a galaxy similar to the Milky Way \citep{dow92}. The
dashed horizontal line, with log($L_{\rm FIR}/D^2_{\rm
25}$)~=~7.25~\solum\,kpc$^{-2}$, is used as the borderline between
starburst and non-starburst galaxies throughout this paper.}
\label{fig:starburst-def}
\end{figure*}

Here we select $L_{\rm FIR}$ as {\it the} measure of the SFR and define
a starburst galaxy in terms of its SFR surface density. Lacking
high resolution information, we use with the isophotal diameter
$D_{25}$ the ratio $L_{\rm FIR}$/$D^2_{\rm 25}$ to determine the
SFR surface density and classify targets with log~($L_{\rm FIR}$/${D^2_{25}}$)
$\geq$ 7.25~$\solum$~kpc$^{-2}$ as starburst galaxies. This
parameter is plotted as a function of log~$L_{\rm FIR}$ in
Fig.~\ref{fig:starburst-def}, where the boundary between
starburst and non-starburst galaxies is marked by a dashed horizontal
line. The borderline was chosen to ensure that most of the well
known starburst galaxies are properly classified. Galaxies, which
were classified as starbursts in the literature (regardless of the
details of the definition), are marked as stars. While all (U)LIRGs
and most of the well known starburst galaxies are well above the
borderline, there are 19 galaxies that were ``misclassified'' (following our
definition) as non-starbursts and 7 galaxies that were ``misclassified''
as starbursts. NGC~253, a typical starburst galaxy, and IC~342, a
galaxy similar to our Milky Way galaxy, are part of Fig.~\ref{fig:starburst-def}
to ensure that the classification method is correct. Both are
located in the expected zone. The starburst sub-sample, selected as such,
includes 24 classical starbursts, 16 Seyfert composites, 13 starburst
supported LINERs, 3 dwarf starburst galaxies ($M_{\rm B}$~$>$~--18;
e.g., NGC~1569 ), and all 28 (U)LIRGs, or in total 77 galaxies
(some of these galaxies have more than two assignments).

Characterized by FIR luminosity and nuclear activity, the entire
sample consists of 4 ULIRGs, 24 LIRGs, 45 Seyferts, 45 LINERs, 49
starbursts neither being ULIRGs nor LIRGs, and 11 ``normal'' galaxies.
Note that one object may be part of more than one sub-sample. The sample
classification is presented in more detail in Table~\ref{tbl:type}.

The sub-sample of Seyferts is severely biased to Seyfert 2 galaxies,
with only 7 galaxies classified as Seyfert 1. Although individually not
satisfying the LIRG criterion 10$^{11}$~$\solum$ $\leq$ $L_{\rm FIR}$
$<$ 10$^{12}$~$\solum$, Arp~303~S and N are both classified as LIRGs
since the system as a whole meets the LIRG criterion. There are 16
Seyferts that are also classified as LINERs. Hence there is a total of
74 AGN in our sample. Excluding those overlapping with the starburst
and (U)LIRG sub-samples, there remain 35 galaxies which show ``pure''
AGN activity.

The sample can also be broken down by Hubble types. We observed 42
early-type (including 19 lenticulars, 2 ellipticals, 1 cD, and 20
early-type spirals) and 54 late-type galaxies (5 irregulars and 49
late type spirals), with a Hubble type index of $t$ = 3 \citep[or
Sb in the RC3,][]{dev91,pat03} being used as the dividing line
($t$~$<$3: early type, $t$~$\ge$~3: late type). Because of their
peculiar morphology, the 28 (U)LIRGs of our sample have not been
included. Concerning the presence and the strength of a bar, the
sample covers 42 SA (unbarred), 31 SAB (weakly barred) and 25 SB
(barred) galaxies. It also comprises 11 Virgo cluster galaxies and
8 dwarf galaxies ($M_{\rm B}$~$>$~--18).

\subsubsection{The sample distribution}
\label{section:distribution}

\begin{figure}
\centerline{
\includegraphics[angle=0,width=14.5cm]{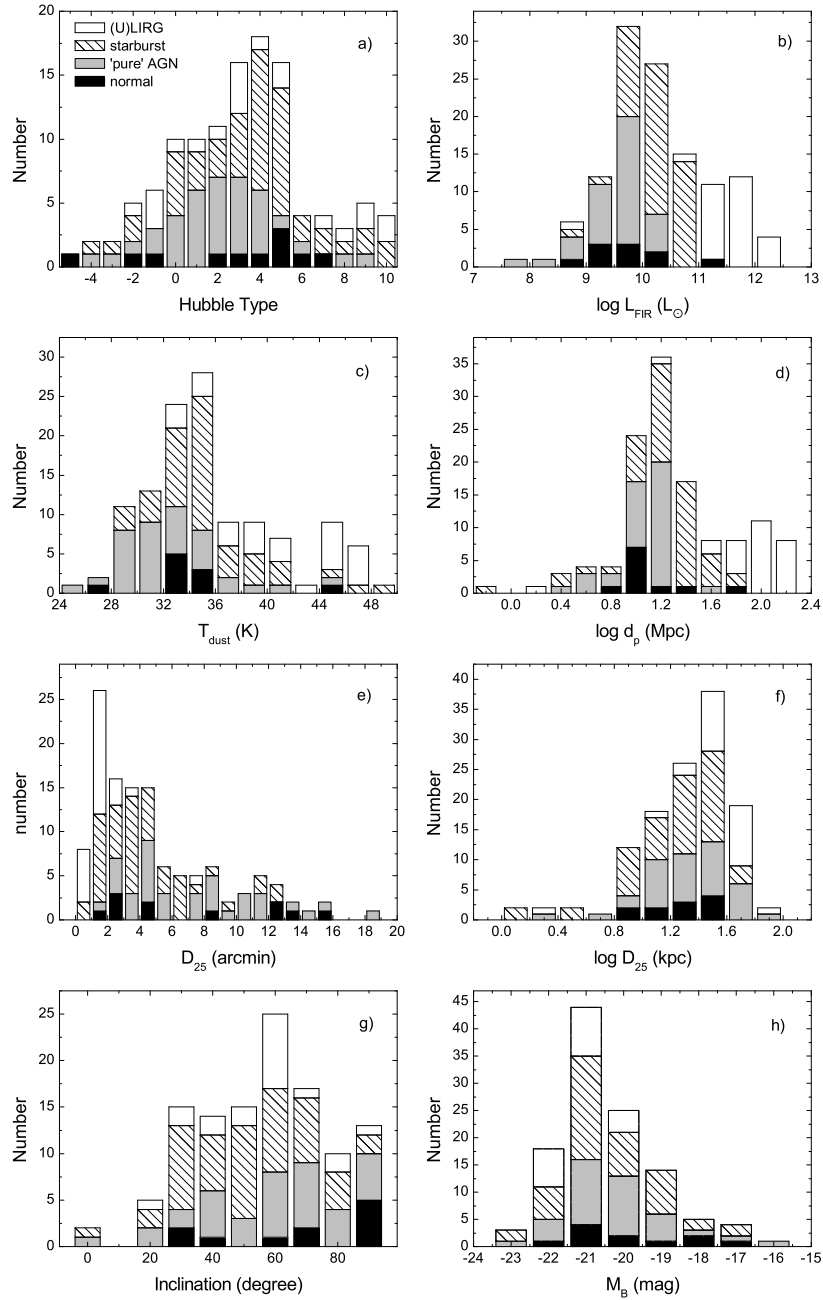}}
\caption[]{Number distributions of the observed galaxy sample: a)
Hubble type \citep[see][]{dev91}, b) far-infrared luminosity, c)
60$\mu$m/100$\mu$m dust color temperature, d) distance,
e) optical angular size, f) linear size, g) inclination,
and h) B-band magnitude of the observed sample.}
\label{fig:distribution}
\end{figure}

In Fig.~\ref{fig:distribution} we present some basic properties of
the entire sample of 125 observed galaxies, i.e. the distribution of
Hubble type, FIR luminosity ($L_{\rm FIR}$), 60$\mu$m/100$\mu$m
dust color temperature ($T_{\rm dust}$), distance ($d_{\rm p}$),
optical angular size ($D_{\rm 25}$ in arcmin), optical linear size
($D_{\rm 25}$ in kpc), inclination angle ($i$), and absolute B-band
magnitude ($M_{\rm B}$). Our sub-samples cover almost all types of
galaxies, with most of them belonging to Hubble types 3 -- 5
\citep[see][]{dev91}, corresponding to the revised (de Vaucouleurs)
morphological types Sb--Sbc--Sc. The far-infrared luminosity of our
sample spans almost 5 orders of magnitude, log($L_{\rm FIR}$/$\solum$)
$\sim$ 7.5--12.4, with a median value of log($L_{\rm FIR}$/$\solum$)
$\sim$ 10.2, which is very close to the total far-infrared luminosity
of the Galaxy \citep[e.g.,][]{bei87}. The dust color temperature
$T_{\rm dust}$, obtained by assuming an emissivity propertional to $\nu$
(see footnote to Table 1), varies between 24\,K and 50\,K, with a peak at about
35\,K. The distance distribution shows a strong peak at 10--20\,Mpc, where
our beam size of 22$''$ (see \S\,\ref{section:obs}) corresponds to a
linear scale of about 1--2\,kpc. These are typical sizes for circumnuclear
starbursts. The optical diameter is in a range between 0.4 to 18.6\,arcmin
on an angular and 1.3 to 78\,kpc on a linear scale, with values of
25--40\,kpc being most typical. About 70\% of our sample galaxies
have an optical diameter ($D_{\rm 25}$) smaller than 5\,arcmin, and the
strong peak at 1--2\,arcmin is due to the merging sequence sub-sample
(see \S2.1). The inclination angle is broadly distributed between
$\sim$30$^\circ$ and 90$^\circ$. There are only few galaxies with
inclinations below $i$ $\sim$30$^{\circ}$. Our sample spans a B-band
absolute magnitude range from --16.4 to --22.6, the majority having
with $M_{\rm B}$ $<$ --20 a high luminosity.

\section{Observations}
\label{section:obs}

All the CO(3--2) observations were conducted with the 10-m Heinrich-Hertz
Telescope (HHT) on Mt. Graham/Arizona with a beamwidth of 22$''$
\footnote{The HHT was operated by the Submillimeter Telescope
Observatory on behalf of Steward Observatory and the Max-Planck-Institut
f{\"u}r Radioastronomie.}. Most of our galaxies were observed during Feb.,
Apr. and Nov. 1999, Jan. and Mar. 2000. The majority of the
merging/interacting galaxies was observed in Mar. 2003, Mar. 2004 and
Mar. 2005. In all cases, the same dual channel 345~GHz SIS
(Superconductor-Insulator-Superconductor) receiver was employed.
Spectral profiles were obtained with two acousto-optical spectrometers
(AOSs), each with 2048 channels (channel spacing $\sim$
480\,kHz, frequency resolution $\sim$ 930\,kHz, corresponding to a
velocity resolution of $\sim$0.8\,km\,s$^{-1}$) and a total bandwidth
of 1\,GHz.

Spectra were taken using a wobbling (2\,Hz) secondary mirror with
beam throws of $\pm$120$''$ to $\pm$240$''$ in azimuth. Scans obtained
with reference positions on either side of the source were coadded
to ensure flat baselines. Receiver temperatures were of order of
170\,K and system temperatures were $\sim$~900\,K on a \TASTAR\
scale, respectively.

Calibration at submillimeter wavelengths is often difficult, especially
for extragalactic observations, and needs to be carefully checked.
The receivers were sensitive to both sidebands. Any imbalance in the
gains of the lower and upper sideband would thus lead to calibration
errors. To account for this, galactic calibration sources (e.g. Orion-KL,
IRC+10216, Sgr~B2, G34.3, and W51, depending on availability at the time
of observation) were observed prior to the target source with the same
receiver tuning setup. Published spectral line survey data in the 345~GHz
band were used for intensity calibrations, e.g. \citet{sch97} for Orion-KL,
\citet{gro94} for IRC+10216, \citet{sut91} for Sgr~B2, \citet{hat98} for
G34.3, and \citet{wan94} for W51. Pointing and focus were carefully
checked before the calibration spectra were taken. Nevertheless, the
absolute calibration error could be as large as $\pm$30\%
(see \S\,\ref{section:consistency}).

After a first order (or second order in very few cases) baseline
subtraction, the antenna temperature \TASTAR\ was converted to
main beam brightness temperature \TMB\  via \TMB\,\,= \TASTAR\
($F_{\rm eff}$/$B_{\rm eff}$) \citep[see,][]{dow89}. The main beam
efficiency, $B_{\rm eff}$, was 0.5 at 345\,GHz, as obtained from
measurements of Saturn, and the forward hemisphere efficiencies,
$F_{\rm eff}$, was 0.9 \citep[see also][]{mao02}.

To reduce as much as possible the number of receiver tunings, we
used the same tuning setup to observe as many galaxies as
possible with similar velocities. Therefore, in some cases the
line is detected well outside the center of the spectrum and
sometimes even reaches the band edge of the backend, in which case
only a zero order baseline subtraction was performed. Additionally,
in a few galaxies like Mrk~273, NGC~6240, IRAS~17208-0014, Arp~220,
and Arp~302N, the full width to zero power of the line is as wide
as $\sim$1000\,km\,s$^{-1}$ \citep[as shown in wide band interferometer
data; e.g.,][]{sco97}, which exceeds the bandwidth of the backend
used for our observations. The intensities in such cases can only
be considered as lower limits, unless some concatenated spectra were
obtained, as in the case of Arp~220 and Arp~302N.

\section{Results}
\label{section:results}

\subsection{Spectra and line parameters}
\label{section:spectra}

Figure~\ref{fig:co-spectra} shows the CO(3--2) spectra (on a \TMB\
scale) towards all detected galaxies. The line parameters or the
upper limits in case of non-detections are given in Table~\ref{tbl:para}.
The spectra have been smoothed to a velocity resolution of
$\sim$3--20\,\kms\ in order to show the emission features more
prominently. Spectra with the velocity integrated intensity
$I_{\rm 32}$ = $I_{\rm CO(3 - 2)}$ = $\int T_{\rm mb}\,{\rm d}v$
larger than three times the r.m.s noise are considered to be detected.
We have determined the integrated line intensity, the radial velocity
and the line width using either Gaussian fits to the lines, or the
moments of the spectra in the case of non-Gaussian line profiles.
Spectra of Arp~220 and Arp~302N were concatenated from two different
velocity setups to cover the full velocity ranges that exceed the
bandwidth of the backend. Of the observed 125 galaxies, 114 were
detected, among which CO(3--2) data of 65 galaxies are reported here
for the first time. For spectra with a signal to noise ratio of less
than 3, an upper limit is derived using $I_{\rm 32}<3\sigma$\,($\Delta$$V_{\rm
10}\,\delta$$v$)$^{1/2}$, where $\sigma$ is the r.m.s noise in \TMB\
for a single channel, and $\Delta$$V$$_{\rm 10}$ represents the full
linewidth taken from the FCRAO CO(1--0) survey results
\citep{you95,ken88} or arbitrarily set to 400~\kms\ if there was
no CO(1--0) data available. $\delta v$ denotes the channel spacing.

The CO(3--2) luminosity, $L_{\rm CO(3-2)}$ in units of K~\kms~pc$^2$, is
calculated within our 22$''$ beam by
\begin{equation}
L_{\rm CO(3-2)} = [\pi/(4ln2)]\,\Theta_{\rm mb}^2\,\,I_{\rm 32}\,\,d_{\rm L}^2\,\,(1+z)^{\rm -3},
\label{equ:luminosity}
\end{equation}
where $\Theta_{\rm mb}$ = 22$''$ = 1.067$\times$10$^{-4}$ rad is the full width
to half maximum (FWHM) main beam size of the HHT at 345 GHz, $d_{\rm L}$  =
$d_{\rm c}$(1+$z$) is the luminosity distance in pc ($d_{\rm p}$: proper distance 
in pc, see footnote to Table), and $z$ = $v_{\rm hel}$/c denotes the redshift. The CO(1--0)
luminosity is derived similarly (see footnotes in Table\,\ref{tbl:para}).
Because of identical beam sizes (22$''$), the averaged intensity ratio between
the $J$=3--2 and 1--0 CO lines, $R_{\rm 31}$ = $I_{\rm CO(3-2)}$/$I_{\rm CO(1-0)}$,
is calculated for galaxies with available IRAM-30m CO(1--0). For galaxies with
CO(1--0) data from other telescopes, upper or lower limits are given for
$R_{\rm 31}$, depending on the CO(1--0) beam size.

\subsection{Detection rates and non-detections}
\label{section:det-rate}

Our CO(3--2) detection rates are 91\% (10/11), 86\% (64/74),
100\% (49/49), and 100\% (28/28) in normal, Seyfert/LINER,
starburst galaxies, and (U)LIRGs, respectively, or about 90\%
in total. Fortuitously, both the total number and the detection
rate, 89\% (39/45), are the same for the Seyferts and
LINERs. Those sample galaxies that are known to host 22~GHz
H$_{\rm 2}$O and/or 18~cm OH masers are all detected in CO(3--2).
For the Virgo Cluster galaxies, dwarf galaxies, early type and
late-type galaxies (see \S\,\ref{section:classification}) the
detection rates are 82\% (9/11), 75\% (6/8), 78\% (18/23) and 94\%
(46/49), respectively.  Note again that we adopted the NED based
classification, which may suffer in some cases from ambiguities.

Eleven of our sample galaxies were not detected in CO(3--2) according
to our detection criteria mentioned above. Their CO(1--0)
intensity is weak on average ($<$\,20\Kkms), with the only
exception of NGC~4438 where a relatively strong ($\sim$70~\Kkms)
CO(1--0) intensity was reported. NGC~4438, together with NGC~2841
and NGC~5866 were, however, observed at poor weather conditions.
NGC~7077 is considered to be a tentative detection, since the
central velocity differs by about 140\,\kms\ from that of the
CO(1--0) line, while its integrated intensity marginally satisfies
the detection criteria with an S/N ratio of 4$\sigma$. One of
our non-detections, NGC~855, a dwarf elliptical, was detected
by \citet{nak07} after a deep integration. The reported intensity
is below our 2$\sigma$ level and therefore well below our detection
limit.

It is interesting to note that, except for one dwarf elliptical
(NGC~855), all the rest of the non-detections are AGN hosts
(either Seyferts or LINERs), and are mostly early type galaxies (7
lenticulars and 3 spirals). This is suggestive of a possible
destruction of molecular gas reservoirs by AGN feedback, and
consequently a suppression of star formation in early type
galaxies \citep[see, e.g.,][]{schaw07}. Except for 3 SAB galaxies,
the non-detections were all obtained from unbarred galaxies.

\subsection{Consistency of the observed CO intensities}
\label{section:consistency}

As already mentioned in \S3, calibration uncertainties may rise
up to $\pm$30\%. Furthermore, with a 22$''$ beam, any shift
$\ga$ 5$''$ could yield significant discrepancies in both
line shape and intensity. This could further increase the
uncertainties of measured absolute intensities and requires a
detailed comparison with data from previous surveys with respect
to both intensity and line shape. The large sample analyzed
here brings us into the unique position to test not only the quality
of our own data but also that of previously studied samples.
The comparison of spectroscopic results is given in the
Appendix and starts with previous measurements also obtained
with a 10-m sized telescope (\S A.1) and continues with the
inclusion of data from the James Clerk Maxwell 15-m telescope
(JCMT, \S A.2).

In general, our results are consistent with published
data. The inconsistencies found for a few individual sources can
be attributed to errors of pointing, calibration and baseline
subtraction (especially for broad spectra) which are difficult to
quantify. A typical error of $\sim$30\% is not unusual even for
millimeter observations.  Therefore, the inconsistencies shown in
the Appendix are within expected ranges, still leaving space for
significant improvements, possibly obtained by mapping the galaxy
cores. Among the sources with large discrepancies in intensities
($>$~50\%), four (NGC~891, NGC~3079, M~83, and Arp~220) have been
reported with both weaker and stronger intensities in the
literature, leaving our intensities close
to the medium values. Among galaxies with published CO(3--2) maps
available, large discrepancies are found for four out of
a total of twelve sources (NGC~891, NGC~2146, NGC~3593, and
NGC~4631). Skewed profiles appear in a few spectra of
Fig.~\ref{fig:co-spectra} and are most likely caused by pointing
errors. Some galaxies show, however, off-centered CO emission,
and the profiles from the nuclear regions are thus not necessarily
symmetric. Considering all these uncertainties and also to keep
the uniformity of the data set, we will exclusively use our data
for the following discussion. Given the large sample, a few
galaxies with relatively large calibration errors should not affect
the overall correlations (see also \S\,\ref{section:correlation}).

\section{Discussion}
\label{section:discussion}

\subsection{The CO(3--2)/(1--0) Line Intensity Ratio}
\label{section:r31-related}

The beam averaged integrated intensity ratio, $R_{\rm 31}$ =
$I_{\rm CO(3-2)}$/$I_{\rm CO(1-0)}$, can serve as an indicator of
the molecular gas excitation since the ratio is sensitive to the
temperature and density of the molecular gas \citep[e.g.,][]{mau99}.
Although the excitation status of the molecular gas cannot be
accurately determined with only two optically thick transitions,
i.e. CO(3--2) and (1--0), we can use $R_{\rm 31}$ to constrain the
molecular gas temperature and density with either Large Velocity
Gradient (LVG) or Photon-Dominated Region (PDR) models. While
it is likely that $R_{31}$ is varying within the region observed
(see, e.g., \citealt{dum01}, who find that CO(3--2) is more
centrally concentrated), our $R_{31}$ values provide a
representative average over the size of the beam.

\subsubsection{Molecular Gas Excitation Traced by $R_{\rm 31}$}

Considering $\Lambda$ = [$n$(CO)/$n$(H$_{2}$)]/(d$v$/d$r$) =
10$^{-4}$ and 10$^{-5}$ (km s$^{-1}$/pc)$^{-1}$, consistent with
\citet{mau99}, one-component LVG calculations
\citep[c.f.][]{sco74,hen80,mao00} with the recent collision rates of
\citet{flo01} show that $R_{\rm 31}$~$\geq$~1 corresponds to
$T_{\rm k}$~$\ga$~60~K and an H$_{2}$ density of $n$(H$_{\rm 2}$)
$\ga$ 10$^{3.5}$~\percc. Both of these values are larger than
those given in \citet{mau99}, where old collision rates were
used. For the extreme case of $R_{\rm 31}$ = 1.9, a gas
temperature of at least $\sim$200~K is required. A ratio of
$R_{31}$~=~0.2 would instead indicate  $n$(H$_2$)~$<$~300~\percc\
for $\Lambda$ = 10$^{-4}$~(km s$^{-1}$/pc)$^{-1}$, or
$n$(H$_2$)~$<$~2500~\percc\ for $\Lambda$ =
10$^{-5}$~(km s$^{-1}$/pc)$^{-1}$ if $T_{\rm k}$ = 20 -- 60~K.

The thermal budget of the interstellar molecular gas in starburst
regions can be described predominantly in terms of a PDR scenario
\citep[see, e.g.,][]{hol97,mao00,sch07}. If we take a typical strength
of the incident far-ultraviolet (FUV) radiation field, $G_{\rm
0}\sim$10$^{2.8-3.9}$ \citep[in units of the local galactic flux,
1.6$\times$10$^{-3}$~erg~s$^{-1}$~cm$^{-2}$, c.f.][and references
therein]{mao00}, as for the starburst in M82, the standard PDR
model \citep{kau99} results in  $n$(H$_{\rm 2}$) =
10$^{3.9-4.9}$~\percc\ and cloud surface temperatures of 300 --
600~K for $R_{\rm 31}$ = 1.0 -- 1.6.

Among our 114 galaxies with detected CO(3--2) emission, 68 have
published IRAM-30m CO(1--0) data. These are the best candidates
for a comparison (c.f. \S\,4.1) because of matching beam sizes (22$''$).
Seven of these galaxies (marked with a superscript $\dagger$ in
Col.(10) of Table~\ref{tbl:para}) have been excluded because of
a positional discrepancy by $\geq$~5$''$ (for the coordinates
used by us, see Table~\ref{tbl:basic}). Therefore, there remain
61 sources for our analysis. For those galaxies with no available
IRAM-30m CO(1--0) data but with measurements from other telescopes
(i.e. NRAO-12m, FCRAO-14m, SEST-15m, Onsala-20m and NRO-45m), upper
or lower limits were estimated. The results are listed in
Table~\ref{tbl:para}.

Figure~\ref{fig:N-R31} shows the distribution of the $R_{\rm 31}$
ratio of the 61 sources. The distribution has a prominent peak
around 0.5, which is the typical value in the spiral arms of the
Galactic disk \citep[e.g.,][]{oka07}, and an additional minor excess
at about 1.5, which is a sign of highly excited gas as found in the
Central Molecular Zone (CMZ) of the Galaxy \citep[for the Sgr A region
see, e.g.,][]{oka07}. The values of $R_{\rm 31}$ range widely from
0.2 to 1.9 with a mean of 0.81~$\pm$~0.06 (cf.
\citealt{mau99,yao03}), the error being the standard error of the
mean. The $R_{31}$ distribution is deviating from a normal one,
so that (following Chebyshev's inequality) probabilities within
a given range of standard deviations are expected to be moderately
lower than those in the case of a normal distribution.

There are 18 sources with $R_{\rm 31}$~$>$~1, indicative of very
high excitation combined with low optical depth. As expected,
these are mostly starbursts or (U)LIRGs, since such high $R_{\rm 31}$
ratio gas may arise predominantly from UV-irradiated surfaces of
molecular clouds or shocked regions, possibly generated by the interaction
with supernova expansion waves \citep[][and references therein]{oka07},
which are fairly common in starburst regions.

NGC~3310, a starburst galaxy with an exceptionally high
CO(2--1)/(1--0) intensity ratio (2.6$\pm$0.6) as found by
\citet{bra92}, shows also the highest $R_{\rm 31}$ ratio
(1.9$\pm$0.52), confirming the peculiar physical conditions of
the molecular gas in its central region. Enhanced massive star
formation triggered by a recent merging event with a gas-rich
galaxy \citep[e.g][]{bal81,kre01} is most likely responsible
for this peculiar value.

Molecular gas with a high $R_{\rm 31}$ ratio does not {\it
always} need high excitation and has also been found in the Galactic
interarm regions where low density gas dominates \citep{oka07}.
However, such interarm regions should not dominate the overall
CO emission of a spiral galaxy. Alternatively, a warm opaque
cloud veiled by a cool foreground layer of low density, which absorbs
the CO(1--0) but not the CO(3--2) emission, could also raise the
$R_{\rm 31}$ ratio to an exceptional level without participation
of highly excited gas of low opacity (see below).

Table~\ref{tbl:ratios} summarizes the results related to $R_{\rm 31}$
for the different galaxy types outlined in \S\,2.1. In the
following, we will discuss which properties of the galaxies
observed may most efficiently affect the determined $R_{\rm 31}$
ratios.

\subsubsection{Correlations between $R_{\rm 31}$ and galaxy properties}
\label{section:R31-correlation}

In Fig.~\ref{fig:R31-all}, $R_{\rm 31}$ is shown as a function of Hubble
type, distance (or linear beam size), inclination, FIR luminosity,
60$\mu$m/100$\mu$m dust color temperature, and optical size
($D_{25}$).

1) {\it Hubble type} -- There is no correlation between
$R_{31}$ and Hubble type (Fig.\,\ref{fig:R31-all}a,
but see \citealt{nak07} for elliptical galaxies).
The bulk of the CO emission in the majority of galaxies
arises from the central region, which is largely decoupled
from the Hubble type and the shape of the large scale
disks \citep{ken98}.

2) {\it Projected beam size} --
Fig.~\ref{fig:R31-all}b shows no correlation,
agreeing with \citet{yao03} on a similar analysis for
their sample. Such a lack of correlation also holds within a given
Hubble type of galaxies. It may imply that toward the
nearby galaxies we see exclusively the nuclear region, while in
the more distant more luminous galaxies the central regions become
so dominant that the larger projected beam size is not important
any more.

3) {\it Inclination} -- Inclination may affect the observed
$R_{\rm 31}$, since we tend to include more low excitation gas
from the outer disk into the observing beam for galaxies seen more
edge-on, thus lowering $R_{31}$. Nevertheless,
Fig.~\ref{fig:R31-all}c shows no strong trend between $R_{31}$
and the cosine of the inclination. However, the upper
envelope of the $R_{31}$ distribution as well as the number of
sources with high $R_{31}$ increase with decreasing
inclination. While this agrees with the expected trend,
the correlation is not significant.

4) {\it FIR luminosity and color temperature of the dust} -- The FIR
luminosity and the temperature of the dust are expected to be
correlated with the molecular gas excitation, if dust and gas
are coupled. In Figs.~\ref{fig:R31-all}d and e we therefore correlate
$R_{\rm 31}$ with $L_{\rm FIR}$ and the 60$\mu$m/100$\,u$m dust
color temperature ($T_{\rm dust}$). Although there may be a weak trend
with $T_{\rm dust}$, similar to that mentioned above for the inclination,
no significant correlation is evident. This could be partially
attributed to the fact that we plot the global dust properties against
the rather localized line ratio $R_{\rm 31}$ emphasizing the nuclear
regions.  In Fig.~\ref{fig:R31-all}f we also plot $R_{\rm 31}$
as a function of $L_{\rm FIR}$/$D_{\rm 25}^2$, a measure of the
SFR per unit area, but again there is no convincing
correlation (see also \citealt{yao03}). This can be interpreted in
the sense that star formation is a locally confined activity.

5) {\it Nuclear activity} -- In addition to star formation,
AGN may also provide a source of heating for the surrounding
molecular gas \citep[see, e.g.,][]{mat04}. As summarized in
Table~\ref{tbl:ratios}, the average $R_{\rm 31}$ ratios are
0.65$\pm$0.08 and 0.82$\pm$0.12 for LINERs and Seyferts,
respectively, or 0.78$\pm$0.09 for ``pure'' AGN. Although being
larger than in normal galaxies (0.61$\pm$0.16), these values are
similar to or lower than the mean value of the entire sample
(0.81$\pm$0.06). This may be caused by the fact that AGN activity
is spatially too much confined for the currently achieved
angular resolution.

On the other hand, starbursts and (U)LIRGs do show higher
average $R_{\rm 31}$ ratios (0.89$\pm$0.11 and 0.96$\pm$0.14,
respectively). Among the 18 galaxies with $R_{\rm 31}$ in excess
of unity, 14 belong to starbursts.
This indicates that the presence
of a nuclear starburst is a major reason for a high, beam averaged
$R_{\rm 31}$ value.

6) {\it Bars} -- Bars are expected to enhance the gas flow toward
the center of galaxies, building up the nuclear gas reservoir to
maintain nuclear activity and affecting the molecular gas
excitation. We find that R$_{\rm 31}$ is higher in barred
SB and SAB galaxies (0.88~$\pm$~0.15 and 0.80~$\pm$~0.11, respectively)
than in non-barred SA galaxies (0.62~$\pm$~0.10). This and the fact that
there are no SB galaxies and only three SAB galaxies among our 11
CO(3--2) non-detections, indicates that the presence of bars can
not only enhance the CO gas excitation, but is also
increasing the central molecular gas reservoir.

7) {\it The merging sequence} -- Most of the galaxies in the
merging sequence sub-sample have no corresponding IRAM CO(1--0)
data. However, in one of the intermediate mergers, the
Antennae, we do not see a significant enhancement of CO(3--2),
with $R_{\rm 31}$ being 0.7, 0.3 and 0.8 for NGC~4038,
NGC~4039 and the overlap region, while the mean ratio of the
``late'' mergers NGC~1614, Mrk~231, Arp~220, and IRAS~17208--0014
is 1.0$\pm$0.2. This is consistent with the result of
\cite{lee10} who find a weak correlation based on CO(3--2) data
from the JCMT.

In summary, while there is significant dispersion between individual
objects of a given type, targets hosting an AGN may show higher
$R_{\rm 31}$ values than ``normal'' galaxies, while $R_{31}$ ratios are
highest in starburst and barred galaxies. Surprisingly, significant
correlations of $R_{31}$ with other galaxy properties are not found.

\subsection{FIR-to-CO luminosity correlations and star formation laws}
\label{section:discussion-corr}

The FIR luminosity (as a measure of the SFR in galaxies) is
correlated with the CO luminosity (a measure of the total
molecular content). Based on the scaling relation SFR $\propto$
$n_{\rm gas}^{\rm N}$, this can be expressed in terms of the
so-called Kennicutt-Schmidt law \citep[hereafter KS law;][]{sch59,ken98},
which connects the SFR per area with the total gas surface density, i.e.
$\Sigma_{\rm SFR}\sim\Sigma^{N}_{\rm gas}$, with $N$ = 1.4$\pm$0.15.
The empirical evaluation of the molecular KS law is, however,
tracer-dependent, and molecular line tracers with different critical
densities may give different correlations with sub-linear ($s$$<$1),
linear ($s$=1), or super-linear ($s$$>$1) correlation indices,
which differ from $N$ \citep[e.g,][]{ken98,gs04a,gs04b,baa08,bus08}.

Recent models to interpret the various observed correlation indices
in a uniform way \citep{kru07,nar08b} assume that $s$ depends on
the cloud's gas mass above the critical density of the observed molecular
transition. The super-linear SFR--$L_{\rm CO(1-0)}$ correlation is caused
by the fact that the gas density is on average higher than the CO(1--0)
line's critical density, thermalizing most of the CO(1--0) emission.
The sub-linear SFR--$L_{\rm HCN(3-2)}$ correlation \citep{bus08}
is a result of a small fraction of thermal emission as the critical
density of the HCN $J$=3--2 transition is high ($n$(H$_2$)
$\sim$ 10$^7$\,cm$^{-3}$). Observational support for such models is,
however, still fragile given the large uncertainties in (sub)millimeter
molecular line calibration (\S A.1 and A.2) and additional problems
mentioned below. In the following, we will first address some of these
problems and will then derive the corresponding correlations from our data.

\subsubsection{Possible problems in the correlation analysis}
\label{section:cor-problems}

For pointed CO observations, as in our case, a direct power law
correlation of the nuclear CO luminosity with the global FIR luminosity
could be misleading because these two quantities refer to different
spatial scales. To evade such a situation, one should ideally obtain
extended CO maps to measure the entire molecular gas content for a
large sample of galaxies with various Hubble types and FIR luminosity
ranges. Full maps of such a large sample of galaxies in CO(3--2)
are, however, not yet available.

\citet{yao03} employed an alternative way in scaling down the total
FIR luminosity as obtained by the IRAS data to the volume marked by
the angular size of the CO observations. The scaling factor is
determined by the peak-to-total flux density ratio derived from the
corresponding 850~$\mu$m submillimeter continuum images. The basic
assumption for such a technique to be applicable is that the FIR
brightness distribution is similar to that of the 850~$\mu$m
submillimeter continuum. This is, however, most likely not the case
since the FIR emission measured by the IRAS satellite is only sensitive
to warm dust ($T_{\rm dust}$$\ga$~30\,K), while the submillimeter
continuum also traces dust at cooler color temperatures. In most galaxies,
this cooler component dominates. Therefore, the FIR emission is
expected to be much more centrally concentrated than the submillimeter
emission, especially in the case of nuclear starbursts. Indeed,
even the 200~$\mu$m continuum emission has already shown significant
cold dust at large galactocentric radii \citep[e.g.,][]{alt98,kra10}.

Another important aspect is the linear regression fit itself.
Published correlation studies sometimes use $L_{\rm CO}$-$L_{\rm FIR}$
and sometimes $L_{\rm FIR}$-$L_{\rm CO}$. Caution has to be exercised,
however, in comparing these two approaches. Simply taking an inverse
slope (i.e., 1/$s$ instead of $s$) is not appropriate because the standard
linear regression fit assumes that $X$ values are exactly correct, and that
errors or variability only affect the $Y$ values. Hence the regression
of $X$ on $Y$ is different from the regression of $Y$ on $X$. To
determine the line dependent parameter $s$ in analogy to $N$ (\S5.2), line
luminosities and FIR luminosities must be plotted along the x- and y-axes,
respectively. Standard regression fits from figures with inverted axes
(as, e.g., displayed by \citealt{yao03,baa08}) cannot be used.
There is, however, a way to account for errors along both
axes. For this, the publicly available Monte Carlo Markov Chain fitting
packages of \citet{kel07} have been developed.

\subsubsection{The FIR-to-CO correlation}
\label{section:correlation}

In Fig.~\ref{fig:l-l-plot}, we present the correlation between the
nuclear CO line luminosity and the global FIR luminosity $L_{\rm FIR}$.
We performed the fit in both ways, with and without the method
introduced by \citet{kel07}, considering uncertainties along both axes
in the latter case. However, slopes derived with the Kelly packages are
(within a percent) the same as those obtained with the unweighted linear
regression fits. This is perhaps due to the large number and small
intrinsic scatter of our data along both axes in the log-log plots.
In the following we will therefore use the unweighted linear regression
fits.

Intriguingly, both nuclear CO(3--2) and CO(1--0) line luminosities are
tightly correlated to the global FIR luminosity, and our unweighted
linear regression fits result in almost identical slopes slightly
below unity, $s$ = 0.87~$\pm$~0.03 and 0.87~$\pm$~0.05 for CO(3--2) and
CO(1--0), respectively. To account for the fitting problems when exchanging
CO and FIR luminosities (\S\,\ref{section:cor-problems}), we note that
the unweighted linear regression fit for $L_{\rm CO}$-$L_{\rm FIR}$
results in slopes of 0.96~$\pm$~0.04 and 0.94~$\pm$~0.06 for the
CO(3--2) and CO(1--0) luminosities. Both slopes related to CO(3--2)
are smaller than unity, and are not simply reciprocal values.

To avoid a bias by a single discrete data point, IC\,10, with
its by far lowest $L_{\rm CO(3-2)}$ and $L_{\rm FIR}$ values, is
not included in the fits of Fig.\,\ref{fig:l-l-plot}. Fits including
IC\,10 give, however, similar slopes of 0.88$\pm$0.03 and 0.86$\pm$0.04
for CO(3--2) and (1--0), respectively. A fit is also performed replacing
four questionable CO(3--2) intensities by the corresponding peak intensities
obtained from maps (see the discussion of the mapped sources in
\S A.1). The new fit gives the same slope, $s$ = 0.87~$\pm$~0.04.
This is not unexpected because the number of these galaxies is small with
respect to that of the entire sample.

Our slopes are consistent with that derived from 14 local (U)LIRGs
\citep{ion09} but differ from those of \citet{yao03}, who obtained
$L_{\rm CO}$(y-axis)-$L_{\rm FIR}$(x-axis) slopes with scaled FIR 
luminosities (\S5.2.1) well below unity
(0.70 and 0.58 for CO(3--2) and (1--0), respectively; see also
their Fig.~2). However, using instead the uncorrected FIR luminosities,
the slopes become 1.00 and 0.94, respectively, very close to our results.

\subsubsection{On the angular size dependence of the correlation}
\label{section:KS-law}

Instead of following \citet{yao03} and scaling down the FIR
fluxes to the area of the CO observations, we can also test our
results by defining different $D_{\rm 25}$ ranges, assuming
that galaxies with similar optical angular sizes have similar central
(22$''$) to total integrated CO intensity ratios. This assumption is
based on the fact that the diameter of the CO(1--0)-emitting region
$D_{\rm CO}$ is found to be correlated with the optical diameter
$D_{\rm 25}$ as $D_{\rm CO}/D_{\rm 25}\approx 0.5$ \citep{you95}.
The CO(3--2) emission may be more centrally concentrated than the
CO(1--0) emission \citep{dum01} but may follow a similar
relation. In Fig.~\ref{fig:l-l-d25} we compare correlations of
CO-to-FIR luminosities with our CO(3--2) data (panel a)) as well
as the CO(1--0) data from the IRAM-30m telescope (panel b)) and from
the FCRAO-14m/NRAO-12m antennas (panel c)) dividing our sample
galaxies into three groups: 1) $D_{\rm 25}$~$\leq$~2$'$ (filled
circles), 2) 2$'$~$<$~$D_{\rm 25}$~$\leq$~4$'$ (empty triangles)
and 3) 4$'$~$<$~$D_{\rm 25}$~$\leq$~18$'$ (crosses). Corresponding
slopes ($s$) and correlation coefficients ($r$) of linear regression fits
are also given. For a constant $D_{\rm CO}/D_{\rm 25}$ ratio the
resulting slopes should be close to the case of FIR and CO
emission arising from the same region.

The three groups of galaxies show different correlations,
irrespective of the CO transition studied. The compact
galaxies show slopes of order 1.0--1.1, the intermediate sample
is characterized by $s$ $\sim$ 0.8--0.9, and the extended sources
have slopes of 0.6--0.7, all with uncertainties in the range
0.05--0.10. For galaxies in the first group, which includes
most of the (U)LIRGs and some other galaxies at large distance,
most of the CO emission is expected to be covered by the
observing beam. For galaxies in the second and third group,
the CO(3--2) emission should be somewhat or much more extended
but within a given group the scaling factor of the
total-to-central (22$''$) integrated intensities is expected to be
similar.

In summary, and without having to assume equal spatial
distributions of the CO(3--2) and (1--0) emission, we obtain
rising slopes $s$ with increasing compactness of the observed
targets. This can be interpreted in terms of the relationship between
molecular line emission and gas density, anchored by the underlying
KS law \citep{gs04b,kru07,nar08b,bus08} as outlined in \S 5.2. The distant
more compact galaxies, many of them (U)LIRGs, possess in the average
denser gas than the more nearby targets, so that their emission
is more thermalized, yielding a higher slope. Nevertheless,
the roughly linear or sublinear correlations are indicative of a
significant fraction of molecular gas with densities
lower than the critical density of the tracer used. This is
comprehensible in the case of CO(3--2) with its high critical
density of $\sim$10$^5$\,cm$^{-3}$, as the bulk of the nuclear
CO emission should arise from a more diffuse medium ($n$(H$_{\rm
2}$)\,$\sim$\,10$^{3-4}$\,cm$^{-3}$; e.g., \citealt{mau99};
\citealt{mao00}). It is, however, difficult to interpret the
correlation in the same way with CO(1--0) because the density of
the bulk of the nuclear molecular gas may not be lower than
$\sim$10$^{3.5}$\,cm$^{-3}$. Clearly, this deserves further study.
More beam matching CO(1--0) data, more maps providing a measure of
the entire CO(3--2) emission of a galaxy, data from more than two
CO transitions, and infrared data of high angular resolution would
thus be helpful.

\section{Conclusions}

1. CO(3--2) spectra from the central region of a sample of 125
galaxies are presented. With an angular resolution of 22$''$, CO(3--2)
emission is detected in 114 targets. Our survey significantly
increases the number of available CO(3--2) data from galaxies
and provides a reliable data base for future surveys with higher
angular resolution, establishing a bridge to the high $J$ lines
observed toward redshifted targets.

2. The CO(3--2)/(1--0) integrated line intensity ratio $R_{\rm
31}$ varies widely from 0.2 to 1.9. The line ratio appears to
be independent of galaxy properties such as Hubble type and FIR
luminosity and only shows tentative, not significant correlations
with 60$\mu$m/100$\mu$m dust color temperature and inclination angle.

3. To be consistent with common designations but to use at the same time
a clear definition, we have specified the term ``starburst galaxy'' by
the condition log\,[($L_{\rm FIR}$/L$_{\odot}$)/($D_{25}^2$/kpc$^2$)]
$>$ 7.25. 77 of our sample galaxies show this property, which is a measure
of star formation activity per surface area.

4. The average $R_{\rm 31}$ ratios are found to be larger in galaxies
with nuclear activity (AGN and starbursts) or with bars than in those
without. Apparently, these are the galaxies showing enhanced molecular
excitation. Most galaxies with a line ratio of $R_{\rm 31}$~$\geq$~1 are
starbursts. (U)LIRGs have the highest average $R_{\rm 31}$ value, which
may be caused by particularly vigorous activity triggered by galaxy
interaction and merging.

5. The nuclear CO luminosities show a slightly sublinear
correlation with the global FIR luminosity in both the CO(3--2)
and the (1--0) lines. Subdividing our sample into several
bins with different angular sizes to compensate for the different
size of the regions from where CO and FIR emission have been
measured reveals significant differences. Compact and thus
mostly distant luminous galaxies show the largest slopes, possibly
a consequence of relatively high overall molecular densities,
yielding larger fractions of thermalized gas. A similar trend
for CO(1--0) is more difficult to explain, because this
would require densities below 10$^{3.5}$\,cm$^{-3}$.

\section*{acknowledgement}
We wish to thank an anonymous referee for critically reading
the manuscript. We also wish to thank the HHT staff for their
enthusiastic support of the project and for their flexibility in changing
schedules according to variable weather conditions. We also thank
Dr. Jiangshui Zhang for his help in preparing some of the tables and
Dr. B. Kelly for a latest version of his software package. We
acknowledge useful discussions with Drs. M. Dumke and A. Wei{\ss}.
RQM is partly supported by NSFC under grants 10373025 and
10733030.

\section*{Appendix}
\subsection*{A.1. Comparison of CO(3--2) spectra taken with the same
angular resolution}
\label{section:comparison}

\renewcommand{\labelenumi}{\arabic{enumi})}
\begin{enumerate}
\item A comparison of 22 galaxies with the data by
\citet[][HHT-M99]{mau99} shows good agreement within the errors.
Eight sources were re-observed at different positions, four of them
at positions displaced by more than 10$''$ (NGC~3627, NGC~3628,
NGC~6946, and NGC~7541), with the new data showing more symmetric
profiles and stronger intensities. The significant difference
in observed line shapes toward NGC~2146 is likely due to a position
offset of 6$''$, and the flat-topped profile of \citet{mau99} looks
more like a line from the center of the galaxy than our sharply
peaked profile. For most of the other galaxies (NGC~3227, NGC~3351,
NGC~3368, NGC~4414, NGC~4818), our new data show higher quality
profiles although the integrated intensities are quite consistent.

\item  \citet[HHT-D01]{dum01} presented extended CO(3--2) maps
toward nine of our sample galaxies where we can check the pointing
by comparing our line profile with their individual spectra. For
Maffei~2, M~82, M~51, and NGC~6946, \citet{dum01} present spectra,
which are consistent with ours with respect to both lineshape
and intensity. For NGC~3628 and M~83, our profiles and
intensities resemble theirs at the (10$''$,0$''$) and
(--10$''$,0$''$) offsets, respectively, where the line intensities
are about 30\% weaker than the peak intensities at their
(0$''$,0$''$) positions. Relatively large discrepancies exist for
three sources. The line shape of NGC~4631 resembles that given by
\cite{dum01} but their intensity is about twice as high. The other
two are observed at positions slightly different from the nominal
position used by \cite{dum01}. Our position of NGC~891 corresponds
to their (--6$''$,--7$''$) offset position, where their integrated
intensity of about 50 \Kkms\ is tripling our value (17.3 \Kkms).
The position we observed for NGC~2146 is 5$''$ east of their
reference position, but our spectrum looks more like theirs at the
(10$''$,--10$''$) offset position. At this position, their
intensity is about twice as strong as ours. The spectrum by
\citet{mau99} is consistent with \citet{dum01} with respect to
both line shape and intensity, although the nominal positions
differ by about 10$''$. Our observations of all these three source
seem to suffer from large calibration errors, and NGC~2146 may
suffer from an additional pointing error.

\item \citet[HHT-V02]{vil03} covered five early type galaxies of
our sample, with NGC~404 and NGC 4691 being slightly stronger
(by 25\% and 5\%, respectively than our spectra. NGC~855 and
NGC~5666 are non-detections in both data sets (NGC~5666 was a
tentative detection in their paper). Our spectrum of NGC~3593
looks more like that of their (10$''$,0$''$) position which
gives an intensity twice as strong as ours.

\item  \citet[HHT-N05]{nar05} observed three of our sample galaxies.
While their integrated intensities for NGC~3079 and Arp~220 are
about twice as large as our values, IRAS~17208-0014 shows less
than half of the strength we got. A detailed check is not possible,
however, since their given positions (their Table 1) are erroneous.

\item  For the six  common sources also observed by
\citet[CSO-B06]{bay06} with the CSO-10m telescope, we find
consistent results for NGC~3079, NGC 6946 and the Antennae system.
Our intensities of Mrk~231, M83, and Arp 220 are, however, all
about twice as large as theirs. Their CO(3--2) spectrum of
NGC~4736 looks more symmetric than ours and has an intensity,
which is 40\% higher.

\item  Among the five common sources also observed by
\citet[ASTE-K07]{kom07} with the ASTE-10m telescope, four galaxies
(NGC~1068, NGC~1084, NGC~1087 and NGC~7479) were observed at
similar (offsets $\leq$6$''$) positions and show quite consistent
results. The large discrepancy in the case of NGC~157 is due to a
position offset of about one telescope beam (22$''$). Our spectra
show generally better baselines thanks to the backend, which is
twice as wide as theirs.

\end{enumerate}

To evaluate consistencies on a quantitative basis, we define a
relative intensity deviation as log($I'/I_{\rm 0}$),
where $I'$ is the integrated CO(3--2) intensity obtained from the
literature and $I_{\rm 0}$ is from this work.
Figure~\ref{fig:consistency} shows the relative intensity
deviations for galaxies with CO(3--2) data
available in the articles mentioned above.
About 80\% of the data points, excluding those observed at
nominal position offsets $\ga$~5$''$ (open squares in Fig.~\ref{fig:consistency}),
are falling into the $\pm$0.3 dex deviation limits.

\subsection*{A.2. Comparison of CO(3--2) spectra taken with different
angular resolution}
\label{section:comparison-JCMT}

Comparisons with observations at different angular resolution are
not straightforward and need to be treated with utmost caution,
since the molecular gas is rarely smoothly distributed in galaxies,
and any simple scaling could easily become artificial. For the
galaxies with more or less known structure, comparisons of line
profiles and intensities can, however, still be helpful to check
consistency.

In Fig.~\ref{fig:JCMT}, we compare our results with published
CO(3--2) data taken with the JCMT-15m telescope by
\citet[JCMT-Y03]{yao03} and \citet[JCMT-W08]{wil08}. Two straight
lines denote the theoretical relationship of intensities obtained
with the JCMT and HHT assuming point-like (dashed) and uniformly-extended
(dotted) structures with respect to the observing beams. Most of
the sources are located between these two lines, as expected in
case of well calibrated intensities. In general, the JCMT-15m CO(3--2)
data tend to yield higher intensities, as expected, given their
higher angular resolution. This also holds for M~83 \citep{mur09}
and NGC~3521 and NGC~3627 \citep{war10}, with the former two galaxies
revealing integrated CO(3--2) intensity compatible with our values,
while their CO(3--2) emission peak toward the latter is stronger by a
factor of 2.5.

Toward the interacting Arp~302~N/S system, \citet{yao03} observed a
position in between the pair of nuclei, where the CO emission is weak
and where we only obtained an upper limit (indicated by the arrow
pointing towards the left). This refers to ``Arp~302~center'' in
Tables~\ref{tbl:basic} and \ref{tbl:para}. We also observed this system
at the positions of its two nuclei (i.e., Arp~302~N/S), where emission
is stronger. Sources, where JCMT CO(3--2) intensities were derived from
maps, are labeled by arrows pointing downwards. Our CO(3--2) intensity of
NGC~3690 (or Arp~299 in \citealt{wil08}) is a sum of NGC~3690~A and
B. Differences between the JCMT and our HHT integrated intensities
are most pronounced toward Mrk~848 and NGC~5258. While in the case of
Mrk~848 this may be due to differences in calibration, the results
from NGC~5258 may be caused by an unusual gas morphology
\citep[see also][]{wil08}.

\clearpage


\tabletypesize{\scriptsize}
{\tiny\setlength\tabcolsep{2pt} \begin{longtable}{@{}rl@{}r@{}r@{}r@{}r@{}r@{}r@{}r@{}r@{}r@{}c@{}@{}rl@{}}
\caption{The HHT extragalactic CO(3--2) survey sample and galaxy properties
}\\
\hline
\hline \hline
\multicolumn{1}{c}{No.}              & \multicolumn{1}{c}{SOURCE}        &   \multicolumn{1}{c}{R.A.$_{\rm 2000}$} &
\multicolumn{1}{c}{DEC$_{\rm 2000}$} & \multicolumn{1}{c}{$v_{\rm hel}$} & \multicolumn{1}{c}{$d_{\rm p}$}         &
\multicolumn{1}{c}{$D_{\rm 25}$}     & \multicolumn{1}{c}{$i$}           & \multicolumn{1}{c}{M$_{\rm B}$}         &
\multicolumn{1}{c}{log~$L_{\rm FIR}$} & \multicolumn{1}{c}{$T_{\rm dust}$}   & \multicolumn{3}{c}{Classification}  \\
\cline{12-14}
   & &  \multicolumn{1}{c}{(~$^h$~~$^m$~~$^s$)} & \multicolumn{1}{c}{(~~$^{\circ}$~~$'$~~$''$)} & \multicolumn{1}{c}{(\kms)} & \multicolumn{1}{c}{(Mpc)} & \multicolumn{1}{c}{($'$)}   & \multicolumn{1}{c}{(deg)} & \multicolumn{1}{c}{(mag)}
    &  \multicolumn{1}{c}{(L$_{\rm \odot}$)} &  \multicolumn{1}{c}{(K)} & \multicolumn{3}{c}{}  \\
 \multicolumn{1}{c}{(1)}  & \multicolumn{1}{c}{(2)} & \multicolumn{1}{c}{(3)} & \multicolumn{1}{c}{(4)}  & \multicolumn{1}{c}{(5)}  & \multicolumn{1}{c}{(6)} &  \multicolumn{1}{c}{(7)} &  \multicolumn{1}{c}{(8)} &   \multicolumn{1}{c}{(9)} & \multicolumn{1}{c}{(10)} & \multicolumn{1}{c}{(11)} & \multicolumn{1}{c}{(12)}  & \multicolumn{1}{c}{(13)} & \multicolumn{1}{c}{(14)} \\
\hline
\endfirsthead
\caption[]{(continued)}\\
\hline \hline
\multicolumn{1}{c}{No.}              & \multicolumn{1}{c}{SOURCE}        &   \multicolumn{1}{c}{R.A.$_{\rm 2000}$} &
\multicolumn{1}{c}{DEC$_{\rm 2000}$} & \multicolumn{1}{c}{$v_{\rm hel}$} & \multicolumn{1}{c}{$d_{\rm p}$}         &
\multicolumn{1}{c}{$D_{\rm 25}$}     & \multicolumn{1}{c}{$i$}           & \multicolumn{1}{c}{M$_{\rm B}$}         &
\multicolumn{1}{c}{log~$L_{\rm FIR}$} & \multicolumn{1}{c}{$T_{\rm dust}$}   & \multicolumn{3}{c}{Classification}  \\
\cline{12-14}
   & &  \multicolumn{1}{c}{(~$^h$~~$^m$~~$^s$)} & \multicolumn{1}{c}{(~~$^{\circ}$~~$'$~~$''$)} & \multicolumn{1}{c}{(\kms)} & \multicolumn{1}{c}{(Mpc)} & \multicolumn{1}{c}{($'$)}   & \multicolumn{1}{c}{(deg)} & \multicolumn{1}{c}{(mag)}
    &  \multicolumn{1}{c}{(L$_{\rm \odot}$)} &  \multicolumn{1}{c}{(K)} & \multicolumn{1}{c}{}  & \multicolumn{1}{c}{} & \multicolumn{1}{c}{}  \\
 \multicolumn{1}{c}{(1)}  & \multicolumn{1}{c}{(2)} & \multicolumn{1}{c}{(3)} & \multicolumn{1}{c}{(4)}  & \multicolumn{1}{c}{(5)}  & \multicolumn{1}{c}{(6)} &  \multicolumn{1}{c}{(7)} &  \multicolumn{1}{c}{(8)} &   \multicolumn{1}{c}{(9)} & \multicolumn{1}{c}{(10)} & \multicolumn{1}{c}{(11)} & \multicolumn{1}{c}{(12)}  & \multicolumn{1}{c}{(13)} & \multicolumn{1}{c}{(14)} \\
\hline
\endhead
\hline
\endfoot
\hline
\endlastfoot
1 & IC 10 & 00 20 27.4 & +59 17 15 & --348 & 0.7$^a$ & 6.8 & 31 & \nodata & 7.6\phantom{0} & 32.9$^c$ & S & 9.9 & dIrr IV/BCD \\
2 & NGC 157 & 00 34 45.3 & --08 23 51 & 1652 & 22.6\phantom{0} & 4.2 & 62 & --21.3 & 10.4\phantom{0} & 33.1\phantom{0} & S & 4.0 & SAB(rs)bc HII \\
3 & NGC 404 & 01 09 27.7 & +35 43 08 & --48 & 3.1$^a$ & 2.67 & 0 & --16.3 & 7.8\phantom{0} & 34.4$^c$ & & --2.8 & SA(s)0-: LINER \\
4 & NGC 660 & 01 43 02.4 & +13 38 45 & 850 & 11.6\phantom{0} & 8.3 & 78 & --19.1 & 10.4\phantom{0} & 36.9\phantom{0} & S & 1.3 & SB(s)a pec;HII LINER \\
5 & III ZW 35 & 01 44 30.7 & +17 06 09 & 8225 & 112.0\phantom{0} & 0.4 & \nodata & \nodata & 11.6\phantom{0} & 45.3\phantom{0} & L & \nodata & LIRG Sy2 \\
6 & NGC 855 & 02 14 03.8 & +27 52 38 & 595 & 9.7$^a$ & 2.6 & 90 & --16.9 & 8.5\phantom{0} & 35.6$^c$ & N & --4.6 & E \\
7 & NGC 891 & 02 22 32.5 & +42 20 48 & 528 & 9.8$^a$ & 13.5 & 88 & --20.1 & 10.3\phantom{0} & 32.0\phantom{0} & N & 3.0 & SA(s)b? sp HII \\
8 & NGC 972 & 02 34 13.1 & +29 18 43 & 1543 & 21.1\phantom{0} & 3.3 & 66 & --20.4 & 10.6\phantom{0} & 36.7\phantom{0} & S & 2.0 & Sab HII \\
9 & MAFFEI 2 & 02 41 55.2 & +59 36 11 & --17 & 2.8$^a$ & 1.57 & 0 & \nodata & 9.3\phantom{0} & 32.8$^c$ & S & 4.0 & SAB(rs)bc \\
10 & NGC 1055 & 02 41 45.2 & +00 26 39 & 994 & 13.6\phantom{0} & 7.6 & 63 & --19.6 & 10.1\phantom{0} & 31.3\phantom{0} & & 3.2 & SBb: sp LINER2 \\
11 & NGC 1068 & 02 42 40.8 & --00 00 47 & 1137 & 15.6\phantom{0} & 7.1 & 21 & --21.3 & 11.0\phantom{0} & 41.5\phantom{0} & L & 3.0 & (R)SA(rs)b;Sy1 Sy2 \\
12 & NGC 1084 & 02 46 00.1 & --07 34 38 & 1407 & 19.3\phantom{0} & 3.2 & 46 & --20.5 & 10.5\phantom{0} & 35.1\phantom{0} & S & 4.9 & SA(s)c HII \\
13 & NGC 1087 & 02 46 25.1 & --00 29 53 & 1517 & 20.8\phantom{0} & 3.12 & 33 & --20.5 & 10.2\phantom{0} & 33.4\phantom{0} & S & 5.2 & SAB(rs)c \\
14 & NGC 1275 & 03 19 48.2 & +41 30 42 & 5264 & 71.9\phantom{0} & 2.2 & 58 & --22.5 & 10.9\phantom{0} & 46.3\phantom{0} & S & --2.2 & cD;pec;NLRG Sy2 \\
15 & NGC 1530 & 04 23 26.6 & +75 17 43 & 2461 & 33.7\phantom{0} & 4.6 & 58 & --21.4 & 10.5\phantom{0} & 32.0\phantom{0} & N & 3.1 & SB(rs)b \\
16 & NGC 1569 & 04 30 46.5 & +64 51 01 & --104 & 2.0$^a$ & 3.6 & 65 & --16.6 & 8.6\phantom{0} & 46.6\phantom{0} & S & 9.6 & IBm;Sbrst Sy1 \\
17 & NGC 1614$^p$ & 04 34 00.0 & --08 34 44 & 4778 & 65.2\phantom{0} & 1.3 & 42 & --21.3 & 11.5\phantom{0} & 45.5\phantom{0} & L & 4.9 & SB(s)c pec;HII:Sy2 \\
18 & NGC 1637 & 04 41 28.3 & --02 51 28 & 717 & 12.0$^a$ & 4 & 31 & --18.4 & 9.4\phantom{0} & 33.0\phantom{0} & N & 5.0 & SAB(rs)c \\
19 & NGC 2146 & 06 18 39.6 & +78 21 19 & 893 & 12.2\phantom{0} & 6 & 54 & --21.0 & 10.7\phantom{0} & 41.3\phantom{0} & S & 2.3 & SB(s)ab pec HII \\
20 & NGC 2559 & 08 17 06.1 & --27 27 27 & 1559 & 21.3\phantom{0} & 4.13 & 64 & --21.0 & 10.6\phantom{0} & 32.3$^c$ & S & 4.5 & SB(s)bc pec: \\
21 & NGC 2655 & 08 55 38.8 & +78 13 28 & 1400 & 19.2\phantom{0} & 4.9 & 66 & --21.1 & 9.3\phantom{0} & 29.6$^c$ & & 0.1 & SAB(s)0/a Sy2 \\
22 & NGC 2681 & 08 53 33.0 & +51 18 53 & 692 & 9.5\phantom{0} & 3.6 & 16 & --19.3 & 9.2\phantom{0} & 38.5\phantom{0} & & 0.4 & (R')SAB(rs)0/a Sy3 \\
23 & NGC 2768 & 09 11 37.7 & +60 02 22 & 1373 & 18.8\phantom{0} & 8.1 & 90 & --21.1 & 9.6\phantom{0} & 44.0$^c$ & & --4.3 & S0\_1/2\_ LINER \\
24 & Arp 55$^p$ & 09 15 54.7 & +44 19 49 & 11782 & 160.1\phantom{0} & 1 & 59 & --21.6 & 11.6\phantom{0} & 37.3\phantom{0} & L & 5.2 & LINER;LIRG HII \\
25 & NGC 2782 & 09 17 15.7 & +39 54 14 & 2543 & 34.8\phantom{0} & 3.5 & 45 & --20.8 & 10.4\phantom{0} & 39.2\phantom{0} & S & 1.1 & SAB(rs)a;Sy1 Sbrst \\
26 & NGC 2841 & 09 22 02.7 & +50 58 36 & 638 & 8.7\phantom{0} & 8.1 & 68 & --20.7 & 9.0\phantom{0} & 25.3$^c$ & & 3.0 & SA(r)b:;LINER Sy1 \\
27 & NGC 2903 & 09 32 09.7 & +21 30 07 & 556 & 7.6\phantom{0} & 12.6 & 56 & --20.8 & 10.0\phantom{0} & 34.2\phantom{0} & N & 4.0 & SB(s)d HII \\
28 & Arp 303S$^p$ & 09 46 20.3 & +03 02 44 & 6002 & 81.9\phantom{0} & 1.06 & 73 & --20.5 & 10.7$^b$ & 32.7$^b$ & L & 2.0 & SB(r)ab:pec \\
29 & Arp 303N$^p$ & 09 46 21.1 & +03 04 17 & 5990 & 81.7\phantom{0} & 1.92 & 77 & --21.6 & 10.8$^b$ & 34.3$^b$ & L & 6.8 & SA(s)cd? pec VLIRG \\
30 & NGC 2985 & 09 50 20.9 & +72 16 44 & 1322 & 18.1\phantom{0} & 4.6 & 38 & --20.7 & 9.9\phantom{0} & 29.5\phantom{0} & & 2.3 & (R')SA(rs)ab LINER \\
31 & NGC 3032 & 09 52 08.2 & +29 14 29 & 1533 & 21.0\phantom{0} & 2 & 26 & --18.8 & 9.4\phantom{0} & 32.9$^c$ & N & --1.8 & SAB(r)0$^\wedge$0$^\wedge$ HII \\
32 & NGC 3034 & 09 55 52.6 & +69 40 47 & 203 & 2.8\phantom{0} & 11.2 & 79 & --18.0 & 10.4\phantom{0} & 48.8\phantom{0} & S & 8.0 & I0;Sbrst HII \\
33 & NGC 3079 & 10 01 58.2 & +55 40 43 & 1116 & 15.3\phantom{0} & 7.9 & 83 & --21.4 & 10.5\phantom{0} & 34.7\phantom{0} & S & 6.5 & SB(s)c;LINER Sy2 \\
34 & NGC 3077 & 10 03 21.1 & +68 44 02 & 14 & 3.8$^a$ & 5.4 & 41 & --17.8 & 8.8\phantom{0} & 37.6\phantom{0} & & 7.9 & I0 pec HII \\
35 & NGC 3110 & 10 04 01.9 & --06 28 29 & 5054 & 69.0\phantom{0} & 1.5 & 65 & --21.8 & 11.2\phantom{0} & 35.2\phantom{0} & L & 3.3 & SB(rs)b pec; HII LIRG \\
36 & NGC 3166 & 10 13 44.9 & +03 25 31 & 1345 & 18.4\phantom{0} & 4.8 & 56 & --20.1 & 9.8\phantom{0} & 33.4\phantom{0} & & 0.2 & SAB(rs)0/a LINER \\
37 & NGC 3169 & 10 14 15.0 & +03 27 57 & 1238 & 16.9\phantom{0} & 4.4 & 57 & --20.2 & 9.9\phantom{0} & 31.2\phantom{0} & & 1.2 & SA(s)a pec LINER \\
38 & NGC 3147 & 10 16 53.6 & +73 24 03 & 2820 & 38.6\phantom{0} & 3.9 & 30 & --22.1 & 10.7\phantom{0} & 28.8\phantom{0} & S & 3.9 & SA(rs)bc Sy2 \\
39 & NGC 3227 & 10 23 30.6 & +19 51 54 & 1157 & 15.8\phantom{0} & 5.4 & 68 & --20.0 & 9.8\phantom{0} & 34.7\phantom{0} & & 1.4 & E2: pec LINER \\
40 & HARO 2 & 10 32 32.0 & +54 24 03 & 1430 & 19.6\phantom{0} & 1.12 & 46 & --18.8 & 9.6\phantom{0} & 44.0$^c$ & S & 9.9 & Im pec HII \\
41 & NGC 3310 & 10 38 46.1 & +53 30 08 & 993 & 13.6\phantom{0} & 3.1 & 31 & --20.0 & 10.2\phantom{0} & 41.9\phantom{0} & S & 4.0 & SAB(r)bc pec HII \\
42 & NGC 3351 & 10 43 57.3 & +11 42 16 & 778 & 10.5$^a$ & 3.07 & 42 & --20.1 & 9.8\phantom{0} & 34.5\phantom{0} & S & 3.0 & SB(r)b;HII Sbrst \\
43 & NGC 3367 & 10 46 34.4 & +13 45 09 & 3040 & 41.6\phantom{0} & 2.02 & 37 & --21.3 & 10.5\phantom{0} & 34.5\phantom{0} & S & 5.2 & SB(rs)c;LINER Sy \\
44 & NGC 3368 & 10 46 45.7 & +11 49 11 & 897 & 11.7$^a$ & 7.6 & 55 & --20.8 & 9.7\phantom{0} & 30.5\phantom{0} & & 1.8 & SAB(rs)ab;Sy LINER \\
45 & NGC 3521 & 11 05 48.9 & --00 02 15 & 801 & 11.0\phantom{0} & 11 & 66 & --21.0 & 10.3\phantom{0} & 32.5\phantom{0} & & 4.0 & SAB(rs)bc;HII LINER \\
46 & NGC 3556 & 11 11 31.8 & +55 40 15 & 699 & 9.6\phantom{0} & 8.7 & 68 & --20.6 & 9.9\phantom{0} & 33.1\phantom{0} & N & 6.0 & SB(s)cd HII \\
47 & NGC 3593 & 11 14 37.1 & +12 49 03 & 628 & 8.6\phantom{0} & 1.46 & 75 & --18.2 & 9.6\phantom{0} & 35.3\phantom{0} & S & --0.4 & SA(s)0/a;HII Sy2 \\
48 & NGC 3627 & 11 20 15.4 & +12 59 42 & 727 & 12.6$^a$ & 9.1 & 57 & --21.1 & 10.5\phantom{0} & 34.7\phantom{0} & S & 3.0 & SAB(s)b;LINER Sy2 \\
 & NGC 3627A & 11 20 15.0 & +12 59 30 & & \nodata & & & --21.1 & \nodata & \nodata & & & \\
49 & NGC 3628 & 11 20 17.0 & +13 35 20 & 843 & 11.5\phantom{0} & 14.8 & 79 & --21.3 & 10.3\phantom{0} & 35.5\phantom{0} & & 3.1 & SAb pec sp;HII LINER \\
50 & NGC 3642 & 11 22 18.4 & +59 04 34 & 1588 & 21.7\phantom{0} & 1.76 & 32 & --20.5 & 9.4\phantom{0} & 30.6$^c$ & S & 4.0 & SA(r)bc: LINER Sy3 \\
51 & NGC 3682 & 11 27 42.7 & +66 35 25 & 1515 & 20.7\phantom{0} & 1.7 & 52 & --18.8 & 9.6\phantom{0} & 33.6$^c$ & S & 0 & SA(s)0/a:? \\
52 & NGC 3690A$^p$ & 11 28 30.9 & +58 33 44 & 3064 & 41.9\phantom{0} & 2.9 & 44 & --20.1 & 11.6\phantom{0} & 47.3\phantom{0} & L & 8.7 & IBm pec HII \\
53 & NGC 3690B$^p$ & 11 28 33.6 & +58 33 46 & 3121 & 42.7\phantom{0} & 2.9 & 44 & --20.1 & 11.6\phantom{0} & 47.3\phantom{0} & L & 8.7 & SBm? pec HII \\
54 & NGC 3810 & 11 40 58.9 & +11 28 20 & 993 & 13.6\phantom{0} & 1.69 & 48 & --20.0 & 9.9\phantom{0} & 32.1\phantom{0} & S & 5.2 & SA(rs)c HII \\
55 & NGC 3982 & 11 56 28.1 & +55 07 30 & 1109 & 22.0$^a$ & 1.69 & 30 & --19.8 & 10.0\phantom{0} & 33.6\phantom{0} & S & 3.2 & SAB(r)b: Sy2 \\
56 & IC 750 & 11 58 51.4 & +42 43 24 & 701 & 9.6\phantom{0} & 2.48 & 66 & --18.1 & \nodata\phantom{0} & \nodata$^c$ & N & 2.2 & Sab: sp \\
57 & NGC 4038$^p$ & 12 01 53.0 & --18 52 03 & 1642 & 22.0$^a$ & 5.2 & 52 & --21.3 & \nodata\phantom{0} & \nodata\phantom{0} & S & 8.9 & SB(s)m pec \\
 & NGC 4038/9 & 12 01 55.1 & --18 53 00 & & 22.0$^a$ & \nodata & \nodata & \nodata & 10.8$^b$ & 36.3$^b$ & \nodata & \nodata & (the overlap region) \\
58 & NGC 4039$^p$ & 12 01 53.6 & --18 53 11 & 1641 & 22.0$^a$ & 3.1 & 71 & --21.3 & \nodata\phantom{0} & \nodata\phantom{0} & S & 8.9 & SA(s)m pecLINERSbrst \\
59 & NGC 4102 & 12 06 22.6 & +52 42 39 & 846 & 11.6\phantom{0} & 2.7 & 58 & --19.2 & 10.2\phantom{0} & 39.2\phantom{0} & S & 3.0 & SAB(s)b?;HII LINER \\
60 & NGC 4138 & 12 09 30.7 & +43 41 16 & 888 & 12.2\phantom{0} & 2.6 & 64 & --18.7 & \nodata\phantom{0} & \nodata\phantom{0} & & --0.8 & SA(r)0+ Sy1.9 \\
61 & NGC 4192$^{\rm v}$ & 12 13 48.3 & +14 54 01 & --142 & 16.5$^a$ & 9.8 & 78 & --21.5 & 9.9\phantom{0} & 30.6\phantom{0} & & 2.5 & SAB(s)ab;HII Sy \\
62 & NGC 4254$^{\rm v}$ & 12 18 49.5 & +14 25 03 & 2407 & 32.9\phantom{0} & 5.4 & 32 & --22.5 & 10.5\phantom{0} & 32.6\phantom{0} & S & 5.2 & SA(s)c \\
63 & NGC 4258 & 12 18 57.5 & +47 18 14 & 448 & 8.4$^a$ & 18.6 & 72 & --20.7 & 9.8\phantom{0} & 28.8$^c$ & & 4.0 & SAB(s)bc;LINER Sy1.9 \\
64 & NGC 4293$^{\rm v}$ & 12 21 12.9 & +18 22 58 & 893 & 12.2\phantom{0} & 5.6 & 59 & --20.0 & 9.6\phantom{0} & 33.4$^c$ & & 0.3 & (R)SB(s)0/a LINER \\
65 & NGC 4303$^{\rm v}$ & 12 21 54.7 & +04 28 20 & 1566 & 16.5$^a$ & 6.5 & 18 & --21.7 & 10.5\phantom{0} & 34.4\phantom{0} & S & 4.0 & SAB(rs)bc;HII Sy2 \\
66 & NGC 4314 & 12 22 32.0 & +29 53 43 & 963 & 13.2\phantom{0} & 4.2 & 16 & --19.7 & 9.3\phantom{0} & 35.0$^c$ & & 1.0 & SB(rs)a LINER \\
67 & NGC 4321$^{\rm v}$ & 12 22 55.2 & +15 49 22 & 1571 & 16.5$^a$ & 7.4 & 30 & --22.0 & 10.3\phantom{0} & 31.9\phantom{0} & & 4.0 & SAB(s)bc;LINER HII \\
68 & NGC 4369 & 12 24 35.9 & +39 22 56 & 1045 & 14.3\phantom{0} & 2.1 & 19 & --18.9 & 9.5\phantom{0} & 35.3\phantom{0} & S & 1.0 & (R)SA(rs)a HII \\
69 & NGC 4395 & 12 25 50.1 & +33 32 45 & 319 & 4.3$^a$ & 13.2 & 90 & --17.2 & 8.1\phantom{0} & 31.1$^c$ & & 8.8 & SA(s)m:;LINER Sy1.8 \\
70 & NGC 4414 & 12 26 27.1 & +31 13 22 & 716 & 21.4$^a$ & 3.6 & 54 & --19.9 & 10.6\phantom{0} & 32.9\phantom{0} & S & 5.1 & SA(rs)c?;HII LINER \\
71 & NGC 4438$^{\rm v}$ & 12 27 45.8 & +13 00 30 & 71 & 16.5$^a$ & 8.5 & 73 & --20.7 & 9.5\phantom{0} & 31.9\phantom{0} & & 0.7 & SA(s)0/a pec: LINER \\
72 & NGC 4457$^{\rm v}$ & 12 28 59.3 & +03 34 16 & 882 & 16.5$^a$ & 2.7 & 35 & --19.0 & 9.6\phantom{0} & 34.8\phantom{0} & S & 0.4 & (R)SAB(s)0/a LINER \\
73 & NGC 4490 & 12 30 36.6 & +41 38 12 & 565 & 9.6$^a$ & 6.3 & 47 & --21.6 & 10.1\phantom{0} & 35.8\phantom{0} & S & 7.0 & SB(s)d pec \\
74 & NGC 4527$^{\rm v}$ & 12 34 08.7 & +02 39 11 & 1736 & 16.5$^a$ & 6.2 & 70 & --21.4 & 10.4\phantom{0} & 34.5\phantom{0} & S & 4.0 & SAB(s)bc;HII LINER \\
75 & NGC 4565 & 12 36 20.8 & +25 59 16 & 1230 & 16.8\phantom{0} & 15.9 & 90 & --22.4 & 10.0\phantom{0} & 27.2\phantom{0} & & 3.2 & SA(s)b? sp Sy3 Sy1.9 \\
76 & NGC 4594 & 12 39 58.8 & --11 37 28 & 1024 & 14.0\phantom{0} & 8.7 & 79 & --22.1 & 9.4\phantom{0} & 28.4$^c$ & & 1.1 & SA(s)a;LINER Sy1.9 \\
77 & NGC 4631 & 12 42 07.6 & +32 32 28 & 606 & 8.3\phantom{0} & 15.5 & 85 & --22.1 & 10.2\phantom{0} & 35.9\phantom{0} & N & 6.6 & SB(s)d \\
78 & NGC 4639$^{\rm v}$ & 12 42 52.4 & +13 15 26 & 1018 & 13.9\phantom{0} & 2.8 & 52 & --19.1 & 9.0\phantom{0} & 29.9$^c$ & & 3.5 & SAB(rs)bc Sy1.8 \\
79 & NGC 4654$^{\rm v}$ & 12 43 56.1 & +13 07 43 & 1046 & 16.5$^a$ & 5.17 & 56 & --20.5 & 10.1\phantom{0} & 31.2\phantom{0} & S & 5.9 & SAB(rs)cd \\
80 & NGC 4666 & 12 45 08.9 & --00 27 38 & 1520 & 20.8\phantom{0} & 4.6 & 70 & --21.0 & 10.5\phantom{0} & 33.3\phantom{0} & S & 4.9 & SABc: LINER \\
81 & NGC 4691 & 12 48 13.4 & --03 19 58 & 1100 & 15.1\phantom{0} & 2.52 & 39 & --19.5 & 9.9\phantom{0} & 38.4\phantom{0} & S & 0.4 & (R)SB(s)0/a pecHII \\
82 & NGC 4710$^{\rm v}$ & 12 49 39.0 & +15 09 55 & 1125 & 16.5$^a$ & 4.9 & 90 & --19.7 & 9.7\phantom{0} & 33.7\phantom{0} & N & --0.8 & SA(r)0+? sp HII \\
83 & NGC 4736 & 12 50 53.5 & +41 07 10 & 308 & 4.7$^a$ & 11.2 & 35 & --19.9 & 9.6\phantom{0} & 37.4\phantom{0} & & 2.4 & (R)SA(r)ab;Sy2 LINER \\
84 & MRK 231 & 12 54 04.8 & +57 08 38 & 12642 & 171.7\phantom{0} & 1.3 & 53 & --22.3 & 12.3\phantom{0} & 47.8\phantom{0} & U & 5.0 & SA(rs)c? pec Sy1 \\
85 & NGC 4818 & 12 56 48.8 & --08 31 26 & 1065 & 14.6\phantom{0} & 4.3 & 90 & --19.6 & 10.0\phantom{0} & 41.3\phantom{0} & S & 2.0 & SAB(rs)ab pec: Sbrst \\
86 & NGC 4826 & 12 56 44.2 & +21 41 05 & 408 & 7.5$^a$ & 10 & 60 & --20.5 & 9.8\phantom{0} & 33.8\phantom{0} & & 2.4 & (R)SA(rs)ab;HII Sy2 \\
87 & NGC 4941 & 13 04 12.9 & --05 33 07 & 1108 & 15.1\phantom{0} & 3.6 & 36 & --19.3 & 9.0\phantom{0} & 30.4$^c$ & & 2.4 & (R)SAB(r)ab: Sy2 \\
88 & NGC 5033 & 13 13 27.4 & +36 35 39 & 875 & 12.0\phantom{0} & 10.7 & 66 & --20.8 & 9.9\phantom{0} & 30.2\phantom{0} & & 5.1 & SA(s)c Sy1.9 \\
89 & NGC 5055 & 13 15 49.1 & +42 02 06 & 504 & 10.3$^a$ & 12.6 & 56 & --21.1 & 10.2\phantom{0} & 29.1\phantom{0} & S & 4.0 & SA(rs)bc HII/LINER \\
90 & UGC 8335A$^p$ & 13 15 30.8 & +62 07 45 & 9230 & 125.6\phantom{0} & 0.85 & \nodata & \nodata & \nodata\phantom{0} & \nodata\phantom{0} & L & \nodata & SC; LINER HII \\
91 & UGC 8335B$^p$ & 13 15 35.0 & +62 07 29 & 9313 & 126.8\phantom{0} & 1.47 & \nodata & \nodata & 11.6$^b$ & 44.4$^b$ & L & \nodata & SC HII \\
92 & Arp 193$^p$ & 13 20 35.3 & +34 08 25 & 6985 & 95.2\phantom{0} & 1.5 & 62 & --20.5 & 11.6\phantom{0} & 40.0\phantom{0} & L & 9.9 & Im: pec;HII LINER \\
93 & NGC 5194 & 13 29 52.5 & +47 11 53 & 463 & 7.1$^a$ & 11.2 & 30 & --20.4 & 10.1\phantom{0} & 33.5\phantom{0} & S & 4.0 & SA(s)bc pec;HIISy2.5 \\
94 & M 83 & 13 37 00.7 & --29 51 59 & 513 & 4.6$^a$ & 12.9 & 46 & \nodata & 10.2\phantom{0} & 35.3\phantom{0} & S & 5.0 & SAB(s)c;HII Sbrst \\
95 & NGC 5256$^p$ & 13 38 17.5 & +48 16 37 & 8353 & 113.8\phantom{0} & 1.22 & & \nodata & 11.4\phantom{0} & 40.4\phantom{0} & L & \nodata & Pec;Sy2;LIRG Sbrst \\
96 & NGC 5257$^p$ & 13 39 52.9 & +00 50 24 & 6798 & 92.7\phantom{0} & 1.8 & 62 & --21.8 & 11.3$^b$ & 37.4$^b$ & L & 3.1 & SAB(s)b pec;HII LIRG \\
97 & NGC 5258$^p$ & 13 39 57.7 & +00 49 51 & 6757 & 92.1\phantom{0} & 1.7 & 34 & --21.3 & 11.0$^b$ & 36.1$^b$ & L & 3.1 & SA(s)b pec;HII LINER \\
98 & NGC 5273 & 13 42 08.3 & +35 39 15 & 1064 & 14.6\phantom{0} & 2.8 & 57 & --18.8 & 8.8\phantom{0} & 40.4$^c$ & & --1.9 & SA(s)0$^\wedge$0$^\wedge$ Sy1.9 \\
99 & MRK 273 & 13 42 51.6 & +56 08 13 & 11326 & 154.6\phantom{0} & 0.72 & 64 & --20.8 & 12.1\phantom{0} & 47.0\phantom{0} & U & \nodata & Ring galaxy;Sy2 LINER \\
100 & NGC 5347 & 13 53 17.8 & +33 29 27 & 2335 & 31.9\phantom{0} & 1.7 & 45 & --19.6 & 9.6\phantom{0} & 35.9$^c$ & & 2.0 & (R')SB(rs)ab Sy2 \\
101 & NGC 5666 & 14 33 09.5 & +10 30 37 & 2221 & 30.4\phantom{0} & 0.95 & 37 & --19.2 & 9.7\phantom{0} & 35.9$^c$ & S & 6.4 & S? \\
102 & Arp 302S$^p$ & 14 57 00.3 & +24 36 25 & 10029 & 136.4\phantom{0} & 0.6 & \nodata & \nodata & $<$11.0$^b$ & \nodata\phantom{0} & L & \nodata & SC HII \\
    & Arp 302 center & 14 57 00.5 & +24 36 44 & 10103 & 137.4\phantom{0} & \nodata & \nodata & \nodata & \nodata & \nodata\phantom{0} & \nodata & \nodata & DBL SYS \\
103 & Arp 302N$^p$ & 14 57 00.7 & +24 37 03 & 10094 & 137.3\phantom{0} & 0.9 & \nodata & \nodata & 11.6$^b$ & 34.0$^b$ & L & \nodata & (Sb);HII LINER \\
104 & NGC 5866 & 15 06 30.2 & +55 45 46 & 672 & 9.2\phantom{0} & 4.7 & 86 & --19.9 & 9.2\phantom{0} & 29.9\phantom{0} & & --1.2 & S0\_3 HII/LINER \\
105 & NGC 5907 & 15 15 52.9 & +56 19 33 & 667 & 9.1\phantom{0} & 12.77 & 87 & --20.9 & 9.5\phantom{0} & 27.8\phantom{0} & N & 5.4 & SA(s)c: sp HII: \\
106 & Mrk 848 & 15 18 05.9 & +42 44 53 & 12049 & 163.7\phantom{0} & 0.9 & 90 & --20.4 & 11.8\phantom{0} & 44.7\phantom{0} & L & --1.7 & S0? pec HII \\
107 & NGC 5953 & 15 34 32.3 & +15 11 42 & 1965 & 26.9\phantom{0} & 1.6 & 44 & --19.6 & 10.3\phantom{0} & 37.4\phantom{0} & S & 0.2 & SAa: pec;LINER;Sy2 \\
108 & Arp 220 & 15 34 57.2 & +23 30 12 & 5434 & 74.2\phantom{0} & 1.5 & 57 & --21.0 & 12.1\phantom{0} & 44.7\phantom{0} & U & 8.4 & S?;LINER;HII Sy2 \\
109 & NGC 6240 & 16 52 58.8 & +02 24 04 & 7339 & 100.0\phantom{0} & 2.1 & 82 & --21.5 & 11.7\phantom{0} & 43.9\phantom{0} & L & --0.2 & I0: pec;LINER Sy2 \\
110 & 17208-0014 & 17 23 22.3 & --00 17 02 & 12834 & 174.3\phantom{0} & 0.4 & 50 & --20.2 & 12.4\phantom{0} & 44.5\phantom{0} & U & 3.8 & Sbrst HII \\
111 & Arp 293$^p$ & 16 58 30.6 & +58 56 19 & 5600 & 76.4\phantom{0} & \nodata & \nodata & \nodata & 11.2\phantom{0} & 32.4\phantom{0} & L & \nodata & GPair \\
112 & NGC 6524 & 17 59 14.9 & +45 53 17 & 5698 & 77.8\phantom{0} & 1.3 & 69 & --20.9 & 10.8\phantom{0} & 34.4$^c$ & S & --2.8 & S0\-: \\
113 & NGC 6670B$^p$ & 18 33 34.1 & +59 53 21 & 8428 & 114.8\phantom{0} & \nodata & \nodata & --20.1 & 11.5\phantom{0} & 38.4\phantom{0} & L & --1.1 & HII \\
 & NGC 6670$^p$ & 18 33 35.1 & +59 53 21 & 8650 & 117.8\phantom{0} & 1 & 78 & \nodata & 11.5\phantom{0} & 38.4\phantom{0} & & --1.1 & TRP SYS \\
114 & NGC 6670A$^p$ & 18 33 37.7 & +59 53 22 & 8719 & 118.7\phantom{0} & \nodata & \nodata & \nodata & 11.5\phantom{0} & 38.4\phantom{0} & L & --1.1 & S HII \\
115 & NGC 6814 & 19 42 40.6 & --10 19 24 & 1563 & 21.4\phantom{0} & 3 & 86 & --21.3 & 10.0\phantom{0} & 30.5\phantom{0} & S & 4.0 & SAB(rs)bc Sy1.5 \\
116 & NGC 6946 & 20 34 51.9 & +60 09 15 & 48 & 5.5$^a$ & 11.5 & 31 & --20.8 & 9.7\phantom{0} & 33.7\phantom{0} & & 5.9 & SAB(rs)cd;Sy2 HII \\
117 & NGC 7013 & 21 03 33.1 & +29 53 47 & 779 & 10.7\phantom{0} & 4 & 90 & --19.5 & 8.9\phantom{0} & 28.9$^c$ & & 0.5 & SA(r)0/a LINER \\
118 & NGC 7077 & 21 29 59.6 & +02 24 51 & 1152 & 15.8\phantom{0} & 0.8 & 79 & --16.8 & 8.5\phantom{0} & 29.5$^c$ & S & --3.9 & BCD/E HII \\
119 & NGC 7217 & 22 07 52.2 & +31 21 35 & 952 & 13.0\phantom{0} & 3.9 & 36 & --20.4 & 9.6\phantom{0} & 29.3\phantom{0} & & 2.5 & (R)SA(r)ab;Sy LINER \\
120 & NGC 7331 & 22 37 03.5 & +34 24 43 & 816 & 15.1$^a$ & 10.5 & 75 & --21.5 & 10.5\phantom{0} & 32.7\phantom{0} & & 3.9 & SA(s)b LINER \\
 & NGC 7331A & 22 37 05.1 & +34 24 36 & & \nodata & \nodata & \nodata & --21.5 & \nodata & \nodata & \nodata & \nodata & \\
121 & NGC 7465 & 23 02 00.8 & +15 57 56 & 1968 & 26.9\phantom{0} & 1.2 & 64 & --19.2 & 10.0\phantom{0} & 39.3\phantom{0} & S & --1.9 & (R')SB(s)0$^\wedge$0$^\wedge$: Sy2 \\
122 & NGC 7469 & 23 03 15.6 & +08 52 26 & 4892 & 66.8\phantom{0} & 1.5 & 30 & --21.7 & 11.5\phantom{0} & 41.8\phantom{0} & L & 1.1 & (R')SAB(rs)a Sy1.2 \\
123 & NGC 7479 & 23 04 56.7 & +12 19 23 & 2381 & 32.6\phantom{0} & 4.1 & 36 & --21.6 & 10.6\phantom{0} & 36.6\phantom{0} & S & 4.3 & SB(s)c;LINER Sy2 \\
124 & NGC 7541 & 23 14 43.7 & +04 32 02 & 2689 & 31.3$^a$ & 3.5 & 75 & --21.5 & 10.7\phantom{0} & 34.6\phantom{0} & S & 4.7 & SB(rs)bc: pec HII \\
125 & NGC 7679 & 23 28 46.8 & +03 30 41 & 5138 & 70.1\phantom{0} & 1.3 & 59 & --21.2 & 11.0\phantom{0} & 39.8\phantom{0} & L & --1.3 & SB0 pec:;HII Sy1 LIRG \\
\label{tbl:basic}
\end{longtable}
{
The columns contain the following information:
Col.(1): The sequence number of the specific source.
Col.(2): Galaxy name;  $^p$: galaxy pair; $^{\rm v}$: Virgo cluster galaxy.
Cols.(3) and (4): Right ascension (R.A.) and declination (DEC) in J2000.0 coordinates.
Col.(5): Heliocentric velocity ($v_{\rm hel}$) from NED.
Col.(6): Galaxy proper distance calculated from $v_{\rm hel}$ using $H_0$ = 73\,km\,s$^{-1}$\,Mpc$^{-1}$ and adopting a flat cosmology with $\Omega_{\rm M}$ = 0.27 and $\Omega_{\rm \Lambda}$ = 0.73.
$^a$: recently measured distances drawn mostly from a crosslink in NED (c.f. NED 1D for references). The distance to Virgo cluster galaxies is set to 16.5 Mpc \citep{mei07}.
Col.(7): Optical diameter ($D_{\rm 25}$) from NED.
Col.(8): Galaxy inclination angle ($i$) from HyperLEDA.
Col.(9): B-band absolute magnitude (M$_{\rm B}$) from HyperLEDA.
Col.(10): FIR luminosity ($L_{\rm FIR}$ = $L$(40--400~$\mu$m)) calculated following the prescription of \citet{mos92}
(see Sec.\ref{section:IRAS}).
$^b$: calculated with infrared fluxes taken from \citet{sur04} where the HIRES processing is adopted allowing for a deconvolution of close galaxy pairs.
$^c$: sources not included in the RBGS \citep{san03};
Col.(11): Dust temperature ($T_{\rm dust}$) derived from the IRAS 60~$\mu$m/100~$\mu$m color assuming
an emissivity that is proportional to the frequency $\nu$.
Col.(12): Galaxy classification from this paper; N: normal; S: starburst; L: LIRG; U: ULIRG. All the rest are ``pure'' AGN (see Sect.\,\ref{section:classification}).
Col.(13): Galaxy type code from HyperLEDA.
Col.(14): Galaxy classification from NED.
}
}

\clearpage
\begin{deluxetable}{lccccc}
\tabletypesize{\scriptsize}
\tablecaption{Galaxy classification of the sample \label{tbl:type}}
\tablehead{ \multicolumn{1}{l}{} & \multicolumn{1}{c}{LINER} &
\multicolumn{1}{c}{Seyfert}      & \multicolumn{1}{c}{Starburst} &
\multicolumn{1}{c}{LIRG} & \multicolumn{1}{c}{ULIRG}
}
\startdata
LINER     & 45 & 16 & 20 &  5 &  2 \\
Seyfert   & 16 & 45 & 27 &  8 &  3 \\
Starburst & 20 & 27 & 77 & 24 &  4 \\
LIRG      &  5 &  8 & 24 & 24 & -- \\
ULIRG     &  2 &  3 &  4 & -- &  4 \\
\enddata
\tablecomments{The diagonal gives the total number of sources of a
        specific class (e.g., there are 45 LINERs). Nondiagonal
        coefficients show the number of targets belonging to at
        least two specific classes (e.g., there are 20 galaxies
        which have been classified both as LINERS and as starburst
        galaxies). For details of the classification, see \S 2.2.2.}
\end{deluxetable}

\clearpage
\tabletypesize{\scriptsize} \tablewidth{0pt}
{\tiny \begin{longtable}{rlrrrrrrrr}
\caption{Observed Quantities of the HHT extragalactic CO $J$ =
3--2 survey
}\\
\hline\hline
\multicolumn{1}{r}{No.} & \multicolumn{1}{l}{SOURCE}  &  \multicolumn{1}{c}{$I_{\rm32}$} &
\multicolumn{1}{c}{$v_{\rm 32}$} & \multicolumn{1}{c}{$\Delta$$v$$_{\rm 32}$} & \multicolumn{1}{c}{$T_{\rm mb}$} &
\multicolumn{1}{c}{log~$L_{\rm CO32}$}  & \multicolumn{1}{c}{log~$L_{\rm CO10}$} & \multicolumn{1}{c}{$R_{\rm 31}$}  & \multicolumn{1}{r}{Ref.}  \\
   & & \multicolumn{1}{c}{(K~km~s$^{-1}$)} & \multicolumn{1}{c}{(km~s$^{-1}$)}  & \multicolumn{1}{c}{(km~s$^{-1}$)}  & \multicolumn{1}{c}{(mK)}
 &  \multicolumn{1}{c}{(K~km~s$^{-1}$~pc$^{2}$)} &  \multicolumn{1}{c}{(K~km~s$^{-1}$~pc$^{2}$)} & & \\
 \multicolumn{1}{c}{(1)}  & \multicolumn{1}{c}{(2)} & \multicolumn{1}{c}{(3)} & \multicolumn{1}{c}{(4)}  & \multicolumn{1}{c}{(5)}  & \multicolumn{1}{c}{(6)} &  \multicolumn{1}{c}{(7)} &  \multicolumn{1}{c}{(8)} & \multicolumn{1}{c}{(9)} & \multicolumn{1}{r}{(10)} \\
\hline
\endfirsthead
\caption[]{(continued)}\\
\hline \hline
\multicolumn{1}{r}{No.} & \multicolumn{1}{l}{SOURCE}  &  \multicolumn{1}{c}{$I_{\rm32}$} &
\multicolumn{1}{c}{$v_{\rm 32}$} & \multicolumn{1}{c}{$\Delta$$v$$_{\rm 32}$} & \multicolumn{1}{c}{$T_{\rm mb}$} &
\multicolumn{1}{c}{log~$L_{\rm CO32}$}  & \multicolumn{1}{c}{log~$L_{\rm CO10}$} & \multicolumn{1}{c}{$R_{\rm 31}$}  & \multicolumn{1}{r}{Ref.}  \\
   & & \multicolumn{1}{c}{(K~km~s$^{-1}$)} & \multicolumn{1}{c}{(km~s$^{-1}$)}  & \multicolumn{1}{c}{(km~s$^{-1}$)}  & \multicolumn{1}{c}{(mK)}
 &  \multicolumn{1}{c}{(K~km~s$^{-1}$~pc$^{2}$)} &  \multicolumn{1}{c}{(K~km~s$^{-1}$~pc$^{2}$)} & & \\
 \multicolumn{1}{c}{(1)}  & \multicolumn{1}{c}{(2)} & \multicolumn{1}{c}{(3)} & \multicolumn{1}{c}{(4)}  & \multicolumn{1}{c}{(5)}  & \multicolumn{1}{c}{(6)} &  \multicolumn{1}{c}{(7)} &  \multicolumn{1}{c}{(8)} & \multicolumn{1}{c}{(9)} & \multicolumn{1}{r}{(10)} \\
\hline
\endhead
\hline
\endfoot
\hline
\endlastfoot
1 & IC 10        & 7.1 $\pm$ 0.3\phantom{0}  & --330 $\pm$\phantom{0} 0  & 14 $\pm$\phantom{0} 1  & 491 $\pm$ 30  & 4.6 $\pm$ 0.02  & 5.1 $\pm$ 0.06  & 0.3 $\pm$ 0.05  & 6$^{a,\dagger}$  \\
2 & NGC 157      & 5.8 $\pm$ 0.6$^*$  & 1733\phantom{00000}    & 123\phantom{00000}    & 43 $\pm$ 10  & 7.6 $\pm$ 0.05  & 8.2 $\pm$ 0.17  & 0.3 $\pm$ 0.10  & 3 $^a$  \\
3 & NGC 404    & 5.5 $\pm$ 0.5\phantom{0}  & --57 $\pm$\phantom{0} 2  & 37 $\pm$\phantom{0} 4  & 140 $\pm$ 22  & 5.8 $\pm$ 0.04  & 6.3 & 0.4 & 12 $^a$  \\
4 & NGC 660  & 71.1 $\pm$ 1.4\phantom{0}  & 843 $\pm$\phantom{0} 3  & 290 $\pm$\phantom{0} 6  & 230 $\pm$ 21  & 8.1 $\pm$ 0.01  & 8.2 $\pm$ 0.01  & 0.7 $\pm$ 0.02  & 3 $^a$  \\
5 & III ZW 35    & 4.4 $\pm$ 0.6$^*$  & 8176\phantom{00000}    & 141\phantom{00000}    & 22 $\pm$\phantom{0} 6  & 8.8 $\pm$ 0.07  & $<$ 9.4    & $<$ 2.0  & 10 $^c$  \\
6 & NGC 855          & $<$ 2.0\phantom{0}  & \nodata\phantom{0000} & ( 100 )    & ( 26 ) & $<$ 6.4   & 6 & $<$   2.5  & 20 $^a$  \\
7 & NGC 891      & 17.3 $\pm$ 1.0\phantom{0}  & 545 $\pm$\phantom{0} 1  & 59 $\pm$\phantom{0} 5  & 274 $\pm$ 40  & 7.3 $\pm$ 0.02  & 8.1 $\pm$ 0.02  & 0.2 $\pm$ 0.01  & 3$^{a,\dagger}$   \\
8 & NGC 972      & 37.0 $\pm$ 0.8\phantom{0}  & 1544 $\pm$\phantom{0} 2  & 225 $\pm$\phantom{0} 5  & 155 $\pm$ 13  & 8.3 $\pm$ 0.01  & $<$ 8.6    & $<$ 2.1  & 1 $^b$  \\
9 & MAFFEI 2     & 157.1 $\pm$ 2.3\phantom{0}  & --23 $\pm$\phantom{0} 1  & 163 $\pm$\phantom{0} 3  & 905 $\pm$ 49  & 7.2 $\pm$ 0.01  & 7.0 $\pm$ 0.01  & 1.6 $\pm$ 0.03  & 8 $^a$  \\
10 & NGC 1055    & 19.6 $\pm$ 0.9\phantom{0}  & 959 $\pm$\phantom{0} 4  & 175 $\pm$ 10  & 105 $\pm$ 18  & 7.7 $\pm$ 0.02  & 8.0 $\pm$ 0.02  & 0.4 $\pm$ 0.03  & 3 $^a$  \\
11 & NGC 1068    & 116.0 $\pm$ 1.4\phantom{0}  & 1158 $\pm$\phantom{0} 1  & 228 $\pm$\phantom{0} 3  & 479 $\pm$ 24  & 8.6 $\pm$ 0.01  & 8.8 $\pm$ 0.01  & 0.5 $\pm$ 0.02  & 3 $^a$  \\
12 & NGC 1084    & 14.9 $\pm$ 0.7\phantom{0}  & 1360 $\pm$\phantom{0} 3  & 144 $\pm$\phantom{0} 7  & 98 $\pm$ 14  & 7.9 $\pm$ 0.02  & 8.2 $\pm$ 0.31  & 0.5 $\pm$ 0.34  & 3 $^a$  \\
13 & NGC 1087    & 7.7 $\pm$ 0.4\phantom{0}  & 1498 $\pm$\phantom{0} 2  & 82 $\pm$\phantom{0} 5  & 88 $\pm$ 11  & 7.6 $\pm$ 0.02  & 7.9 $\pm$ 0.03  & 0.5 $\pm$ 0.04  & 3 $^a$   \\
14 & NGC 1275    & 9.0 $\pm$ 0.8\phantom{0}  & 5276 $\pm$ 12  & 248 $\pm$ 23  & 34 $\pm$ 14  & 8.8 $\pm$ 0.04  & 8.5 & 1.8 & 22 $^a$  \\
15 & NGC 1530    & 30.0 $\pm$ 1.1\phantom{0}  & 2417 $\pm$\phantom{0} 5  & 86 $\pm$\phantom{0} 9  & 109 $\pm$ 34  & 8.6 $\pm$ 0.05  &  8.9  & 0.6 & 29 $^a$  \\
16 & NGC 1569    & 2.9 $\pm$ 0.7\phantom{0}  & --70 $\pm$\phantom{0} 5  & 43 $\pm$ 10  & 64 $\pm$ 20  & 5.2 $\pm$ 0.10  & 5.0 $\pm$ 0.04  & 1.5 $\pm$ 0.37  & 30 $^a$   \\
17 & NGC 1614    & 71.1 $\pm$ 5.9\phantom{0}  & 4729 $\pm$\phantom{0} 9  & 221 $\pm$ 19  & 303 $\pm$ 73  & 9.6 $\pm$ 0.03  & 9.4 $\pm$ 0.01  & 1.6 $\pm$ 0.14  & 5 $^a$   \\
18 & NGC 1637    & 4.9 $\pm$ 0.3\phantom{0}  & 729 $\pm$\phantom{0} 3  & 91 $\pm$\phantom{0} 6  & 51 $\pm$\phantom{0} 8  & 7.0 $\pm$ 0.03  & 7.4 $\pm$ 0.02  & 0.4 $\pm$ 0.03  & 3 $^a$   \\
19 & NGC 2146    & 66.9 $\pm$ 1.5\phantom{0}  & 787 $\pm$\phantom{0} 1  & 124 $\pm$\phantom{0} 3  & 505 $\pm$ 36  & 8.1 $\pm$ 0.01  & 8.4 $\pm$ 0.01  & 0.6 $\pm$ 0.02  & 3$^{a,\dagger}$   \\
20 & NGC 2559    & 37.3 $\pm$ 1.6\phantom{0}  & 1497 $\pm$\phantom{0} 3  & 147 $\pm$\phantom{0} 8  & 238 $\pm$ 33  & 8.3 $\pm$ 0.02  & $<$ 8.9    & $<$ 1.2  & 1 $^b$   \\
21 & NGC 2655    & $<$ 2.6\phantom{0}  & \nodata\phantom{0000} & ( 400 )    & ( 17 ) & $<$ 7.1   & $<$ 7.5   & $<$ 1.8   & 1 $^b$  \\
22 & NGC 2681    & 17.1 $\pm$ 1.7\phantom{0}  & 680 $\pm$\phantom{0} 7  & 137 $\pm$ 14  & 117 $\pm$ 40  & 7.3 $\pm$ 0.04  & 7.5 $\pm$ 0.01  & 0.6 $\pm$ 0.06  & 3 $^a$   \\
23 & NGC 2768    & $<$ 1.5\phantom{0}  & \nodata\phantom{0000} & ( 200 )    & ( 14 ) & $<$ 6.8   & 6.6 & $<$   1.9  & 20 $^a$   \\
24 & Arp 55    & 11.1 $\pm$ 0.7$^*$  & 11907\phantom{00000}    & 305\phantom{00000}    & 31 $\pm$\phantom{0} 7  & 9.5 $\pm$ 0.03  & $<$ 10.0    & $<$ 1.6  & 1 $^b$   \\
25 & NGC 2782    & 31.2 $\pm$ 2.2$^*$  & 2587\phantom{00000}    & 272\phantom{00000}    & 95 $\pm$ 25  & 8.7 $\pm$ 0.01  & $<$ 8.7    & $<$ 1.9    & 1 $^b$   \\
26 & NGC 2841    & $<$ 4.2\phantom{0}  & \nodata\phantom{0000} & ( 200 )    & ( 39 ) & $<$ 6.5   & 6.8 & $<$   0.6  & 3 $^a$   \\
27 & NGC 2903    & 59.6 $\pm$ 3.4$^*$  & 525\phantom{00000}    & 143\phantom{00000}    & 390 $\pm$ 69  & 7.6 $\pm$ 0.03  & 7.8 & 0.6 & 8 $^a$   \\
28 & Arp 303 S    & 18.6 $\pm$ 1.2$^*$  & 6058\phantom{00000}    & 379\phantom{00000}    & 42 $\pm$ 12  & 9.2 $\pm$ 0.02  & \nodata  & \nodata  & \nodata  \\
29 & Arp 303 N    & 37.8 $\pm$ 1.9$^*$  & 6144\phantom{00000}    & 502\phantom{00000}    & 62 $\pm$ 18  & 9.5 $\pm$ 0.03  & \nodata  & \nodata  & \nodata  \\
30 & NGC 2985    & 8.0 $\pm$ 1.3\phantom{0}  & 1226 $\pm$ 15  & 183 $\pm$ 38  & 41 $\pm$ 17  & 7.5 $\pm$ 0.07  & 7.7 $\pm$ 0.01  & 0.7 $\pm$ 0.19  & 3 $^a$  \\
31 & NGC 3032    & 4.1 $\pm$ 0.8\phantom{0}  & 1535 $\pm$ 11  & 130 $\pm$ 31  & 30 $\pm$ 12  & 7.4 $\pm$ 0.08  & 7.7 & 0.4 & 27 $^a$  \\
32 & NGC 3034        & 1056.0 $\pm$ 6.3\phantom{0}  & 228 $\pm$\phantom{0} 1  & 203 $\pm$\phantom{0} 1  & 4890 $\pm$ 166  & 8.0 $\pm$ 0.00  & 7.9 & 1.4 & 8 $^a$   \\
33 & NGC 3079    & 93.4 $\pm$ 5.5$^*$  & 1190\phantom{00000}    & 408\phantom{00000}    & 178 $\pm$ 57  & 8.4 $\pm$ 0.03  & 8.8 $\pm$ 0.01  & 0.4 $\pm$ 0.03  & 25 $^a$   \\
34 & NGC 3077    & 8.0 $\pm$ 0.3\phantom{0}  & 7 $\pm$\phantom{0} 1  & 42 $\pm$\phantom{0} 2  & 181 $\pm$ 12  & 6.2 $\pm$ 0.05  & 6.0 $\pm$ 0.03  & 1.5 $\pm$ 0.10  & 2 $^a$   \\
35 & NGC 3110    & 24.5 $\pm$ 0.8\phantom{0}  & 5071 $\pm$\phantom{0} 5  & 325 $\pm$ 11  & 71 $\pm$\phantom{0} 8  & 9.2 $\pm$ 0.01  & (9.2, 9.8)  & (0.5, 1.6)  & 13 $^{b,d}$   \\
36 & NGC 3166    & $<$ 19.0\phantom{0}  & \nodata\phantom{0000} & ( 300 )    & ( 141 ) & $<$ 7.9   & 8.2 &    0.5  & 12 $^a$   \\
37 & NGC 3169    & 11.0 $\pm$ 1.1$^*$  & 1144\phantom{00000}    & 259\phantom{00000}    & 53 $\pm$ 15  & 7.6 $\pm$ 0.05  & $<$ 8.3    & $<$ 0.4   & 1 $^b$   \\
38 & NGC 3147    & 19.8 $\pm$ 1.0\phantom{0}  & 2904 $\pm$ 12  & 466 $\pm$ 28  & 40 $\pm$\phantom{0} 8  & 8.6 $\pm$ 0.02  &  8.5    &  1.2  & 29 $^{a,\dagger}$  \\
39 & NGC 3227    & 18.1 $\pm$ 1.0\phantom{0}  & 1154 $\pm$\phantom{0} 7  & 244 $\pm$ 16  & 70 $\pm$ 12  & 7.8 $\pm$ 0.03  & 8.2 $\pm$ 0.03  & 0.3 $\pm$ 0.03  & 3 $^a$   \\
40 & HARO 2      & 2.5 $\pm$ 0.3\phantom{0}  & 1452 $\pm$\phantom{0} 5  & 86 $\pm$\phantom{0} 9  & 28 $\pm$\phantom{0} 8  & 7.1 $\pm$ 0.05  & 7.5 $\pm$ 0.03  & 0.4 $\pm$ 0.05  & 21 $^a$   \\
41 & NGC 3310    & 6.9 $\pm$ 0.7\phantom{0}  & 1032 $\pm$\phantom{0} 7  & 135 $\pm$ 13  & 48 $\pm$ 15  & 7.2 $\pm$ 0.04  & 6.9 $\pm$ 0.11  & 1.9 $\pm$ 0.52  & 3 $^a$  \\
42 & NGC 3351    & 26.0 $\pm$ 1.4\phantom{0}  & 783 $\pm$\phantom{0} 6  & 200 $\pm$ 11  & 122 $\pm$ 27  & 7.6 $\pm$ 0.02  & 7.4 $\pm$ 0.05  & 1.5 $\pm$ 0.19  & 3 $^a$  \\
43 & NGC 3367    & 7.5 $\pm$ 1.1\phantom{0}  & 3044 $\pm$\phantom{0} 6  & 86 $\pm$ 15  & 82 $\pm$ 22  & 8.2 $\pm$ 0.08  & $>$ 8.6    & $>$ 0.2  & 13 $^d$   \\
44 & NGC 3368    & 24.6 $\pm$ 1.0\phantom{0}  & 885 $\pm$\phantom{0} 4  & 194 $\pm$ 10  & 119 $\pm$ 19  & 7.6 $\pm$ 0.02  & 7.8 $\pm$ 0.01  & 0.7 $\pm$ 0.04  & 3 $^a$   \\
45 & NGC 3521    & 14.5 $\pm$ 1.4\phantom{0}  & 746 $\pm$\phantom{0} 9  & 178 $\pm$ 19  & 76 $\pm$ 20  & 7.4 $\pm$ 0.04  & (7.1, 8.2)    & $<$0.6~  $>$0.9  & 1$^b$,23$^d$   \\
46 & NGC 3556    & 14.4 $\pm$ 1.8\phantom{0}  & 693 $\pm$\phantom{0} 5  & 94 $\pm$ 17  & 143 $\pm$ 31  & 7.2 $\pm$ 0.06  & (7.0, 7.9)  & (0.9, 1.0)  & 1$^b$,31$^d$   \\
47 & NGC 3593    & 17.3 $\pm$ 1.2$^*$  & 643\phantom{00000}    & 128\phantom{00000}    & 99 $\pm$ 17  & 7.2 $\pm$ 0.02  & 7.8 & 0.3 & 12 $^a$   \\
48 & NGC 3627    & 33.4 $\pm$ 1.3\phantom{0}  & 717 $\pm$\phantom{0} 4  & 211 $\pm$\phantom{0} 9  & 149 $\pm$ 25  & 7.8 $\pm$ 0.02  & 8.1 & 0.6 & 3 $^a$   \\
  & NGC 3627A    & 21.9 $\pm$ 0.9\phantom{0}  & 673 $\pm$\phantom{0} 3  & 142 $\pm$\phantom{0} 8  & 145 $\pm$ 19  & 7.7 $\pm$ 0.02  & 8 & 0.5 & 16 $^a$  \\
49 & NGC 3628    & 140.7 $\pm$ 1.4\phantom{0}  & 870 $\pm$\phantom{0} 1  & 214 $\pm$\phantom{0} 2  & 618 $\pm$ 25  & 8.4 $\pm$ 0.00  & 8.3 $\pm$ 0.02  & 1.1 $\pm$ 0.05  & 3 $^a$   \\
50 & NGC 3642    & 7.4 $\pm$ 1.0\phantom{0}  & 1544 $\pm$ 19  & 235 $\pm$ 39  & 29 $\pm$ 12  & 7.6 $\pm$ 0.06  & \nodata & \nodata & \nodata  \\
51 & NGC 3682    & 12.2 $\pm$ 1.6\phantom{0}  & 1584 $\pm$ 16  & 240 $\pm$ 35  & 48 $\pm$ 16  & 7.8 $\pm$ 0.06  & $<$ 7.6  & $<$ 1.7 & 12 $^e$   \\
52 & NGC 3690A    & 36.4 $\pm$ 2.4\phantom{0}  & 3155 $\pm$\phantom{0} 3  & 125 $\pm$ 12  & 273 $\pm$ 46  & 8.9 $\pm$ 0.03  & 8.8 & 1.2 & 18 $^a$   \\
53 & NGC 3690B    & 48.7 $\pm$ 1.9\phantom{0}  & 3102 $\pm$\phantom{0} 5  & 239 $\pm$ 10  & 191 $\pm$ 33  & 9.1 $\pm$ 0.02  & 8.9 & 1.6 & 18 $^a$   \\
54 & NGC 3810    & 20.5 $\pm$ 1.6\phantom{0}  & 960 $\pm$\phantom{0} 6  & 145 $\pm$ 13  & 133 $\pm$ 24  & 7.7 $\pm$ 0.06  & 7.5 $\pm$ 0.03  & 1.7 $\pm$ 0.18  & 3 $^a$   \\
55 & NGC 3982    & 6.7 $\pm$ 0.7\phantom{0}  & 1185 $\pm$\phantom{0} 3  & 61 $\pm$\phantom{0} 9  & 104 $\pm$ 16  & 7.6 $\pm$ 0.04  & $>$ 7.7    & $>$ 0.4  & 11 $^d$   \\
56 & IC 750      & 37.4 $\pm$ 2.2\phantom{0}  & 691 $\pm$\phantom{0} 6  & 205 $\pm$ 14  & 172 $\pm$ 41  & 7.6 $\pm$ 0.03  & 7.8 $\pm$ 0.03  & 0.6 $\pm$ 0.06  & 3 $^a$   \\
57 & NGC 4038    & 44.0 $\pm$ 1.0\phantom{0}  & 1640 $\pm$\phantom{0} 1  & 87 $\pm$\phantom{0} 2  & 473 $\pm$ 19  & 8.4 $\pm$ 0.03  & 8.6 $\pm$ 0.06  & 0.7 $\pm$ 0.03  & 17 $^a$  \\
  & NGC 4038/9  & 82.6 $\pm$ 2.3$^*$  & 1515\phantom{00000}    & 166\phantom{00000}    & 494 $\pm$ 25  & 8.7 $\pm$ 0.02   & 8.8 $\pm$ 0.01  & 0.8 $\pm$ 0.02  & 17 $^a$    \\
58 & NGC 4039    & 15.2 $\pm$ 1.9$^*$  & 1657\phantom{00000}    & 100\phantom{00000}    & 133 $\pm$ 30  & 8.0 $\pm$ 0.04  & 8.5 $\pm$ 0.09  & 0.3 $\pm$ 0.04  & 17 $^a$   \\
59 & NGC 4102    & 49.0 $\pm$ 3.4$^*$  & 823\phantom{00000}    & 220\phantom{00000}    & 172 $\pm$ 39  & 7.9 $\pm$ 0.02  & 8.2 & 0.6 & 8 $^a$   \\
60 & NGC 4138    & $<$ 5.3\phantom{0}  & \nodata\phantom{0000} & ( 400 )    & ( 34 ) & $<$ 7.0   & $<$ 7.1   & $<$ 3.2  & 1 $^b$   \\
61 & NGC 4192    & 7.0 $\pm$ 1.4\phantom{0}  & --191 $\pm$\phantom{0} 6  & 71 $\pm$ 19  & 93 $\pm$ 14  & 7.4 $\pm$ 0.09  & (7.3, 8.1)  & (0.6, 0.8)  & 14$^b$,26$^c$   \\
62 & NGC 4254    & 24.0 $\pm$ 1.7$^*$  & 2434\phantom{00000}    & 140\phantom{00000}    & 147 $\pm$ 20  & 7.9 $\pm$ 0.03  & (7.9, 8.4) & (0.5, 1.2)  & 14$^b$,23$^c$   \\
63 & NGC 4258    & 64.4 $\pm$ 1.1\phantom{0}  & 407 $\pm$\phantom{0} 3  & 311 $\pm$\phantom{0} 6  & 195 $\pm$ 17  & 7.8 $\pm$ 0.01  & 7.8 & 0.9 & 20 $^a$   \\
64 & NGC 4293    & 20.4 $\pm$ 0.5\phantom{0}  & 933 $\pm$\phantom{0} 2  & 159 $\pm$\phantom{0} 4  & 121 $\pm$ 11  & 7.9 $\pm$ 0.01  & 8.4 & 0.3 & 12 $^a$   \\
65 & NGC 4303    & 33.7 $\pm$ 3.2\phantom{0}  & 1575 $\pm$\phantom{0} 6  & 127 $\pm$ 13  & 250 $\pm$ 52  & 8.1 $\pm$ 0.04  & (8.1, 8.7)  & (0.5, 1.6)  & 14$^b$,23$^c$   \\
66 & NGC 4314    & 37.6 $\pm$ 3.0$^*$  & 1049\phantom{00000}    & 187\phantom{00000}    & 159 $\pm$ 53  & 7.9 $\pm$ 0.03  & 7.7 $\pm$ 0.01  & 1.6 $\pm$ 0.12  & 3 $^a$   \\
67 & NGC 4321    & 56.9 $\pm$ 4.2\phantom{0}  & 1590 $\pm$\phantom{0} 7  & 177 $\pm$ 14  & 302 $\pm$ 84  & 8.3 $\pm$ 0.03  & 8.4 $\pm$ 0.00  & 0.7 $\pm$ 0.05  & 3 $^a$   \\
68 & NGC 4369    & 8.2 $\pm$ 0.6\phantom{0}  & 1035 $\pm$\phantom{0} 2  & 62 $\pm$\phantom{0} 5  & 124 $\pm$ 18  & 7.3 $\pm$ 0.03  & 7.8 & 0.4 & 12 $^a$  \\
69 & NGC 4395    & $<$ 2.8\phantom{0}  & \nodata\phantom{0000} & ( 400 )    & ( 18 ) & $<$ 5.8   & \nodata & \nodata & \nodata  \\
70 & NGC 4414    & 31.2 $\pm$ 0.6\phantom{0}  & 740 $\pm$\phantom{0} 3  & 312 $\pm$\phantom{0} 7  & 94 $\pm$\phantom{0} 9  & 8.3 $\pm$ 0.01  & 8.5 $\pm$ 0.03  & 0.5 $\pm$ 0.04  & 3 $^a$  \\
71 & NGC 4438    & $<$ 11.3\phantom{0}  & \nodata\phantom{0000} & ( 300 )    & ( 84 ) & $<$ 7.6  & 8.4 & $<$   0.2  & 3 $^a$   \\
72 & NGC 4457    & 20.3 $\pm$ 0.7\phantom{0}  & 897 $\pm$\phantom{0} 3  & 168 $\pm$\phantom{0} 7  & 114 $\pm$ 14  & 7.9 $\pm$ 0.02  & $<$ 8.1    & $<$ 2.4  & 1 $^b$   \\
73 & NGC 4490    & 24.4 $\pm$ 3.7\phantom{0}  & 669 $\pm$ 32  & 411 $\pm$ 68  & 56 $\pm$ 28  & 7.5 $\pm$ 0.07  & $<$ 7.3  & $<$ 2.1  & 1 $^b$   \\
74 & NGC 4527    & 37.4 $\pm$ 5.3$^*$  & 1642\phantom{00000}    & 179\phantom{00000}    & 171 $\pm$ 48  & 8.1 $\pm$ 0.06  & $<$ 8.6    & $<$ 1.3   & 1$^b$,31$^d$   \\
75 & NGC 4565    & 8.3 $\pm$ 1.5\phantom{0}  & 1271 $\pm$\phantom{0} 9  & 91 $\pm$ 21  & 86 $\pm$ 28  & 7.5 $\pm$ 0.08  & 7.6 $\pm$ 0.04  & 0.7 $\pm$ 0.14  & 3 $^a$   \\
76 & NGC 4594    & 6.8 $\pm$ 0.8\phantom{0}  & 1086 $\pm$ 17  & 292 $\pm$ 41  & 22 $\pm$\phantom{0} 8  & 7.2 $\pm$ 0.05  & $<$ 7.5  & 0.9 & 1 $^b$   \\
77 & NGC 4631    & 17.7 $\pm$ 0.7\phantom{0}  & 638 $\pm$\phantom{0} 1  & 71 $\pm$\phantom{0} 3  & 233 $\pm$ 22  & 7.2 $\pm$ 0.02  & 7.6 & 0.4 & 24 $^a$ \\
78 & NGC 4639    & $<$ 4.5\phantom{0}  & \nodata\phantom{0000} & ( 400 )    & ( 29 ) & $<$ 7.0   & $<$ 7.5   & \nodata & 14 $^b$   \\
79 & NGC 4654    & 18.9 $\pm$ 1.4\phantom{0}  & 1053 $\pm$\phantom{0} 4  & 95 $\pm$\phantom{0} 7  & 187 $\pm$ 40  & 7.8 $\pm$ 0.03  & 7.9 $\pm$ 0.02  & 0.8 $\pm$ 0.07  & 3 $^a$   \\
80 & NGC 4666    & 36.7 $\pm$ 3.8\phantom{0}  & 1498 $\pm$ 13  & 242 $\pm$ 27  & 142 $\pm$ 45  & 8.3 $\pm$ 0.04  & (8.1, 8.8)    & (0.8, 1.2)  & 1$^b$,31$^d$   \\
81 & NGC 4691    & 12.9 $\pm$ 1.0\phantom{0}  & 1129 $\pm$\phantom{0} 1  & 43 $\pm$\phantom{0} 4  & 282 $\pm$ 50  & 7.6 $\pm$ 0.03  & (7.6, 7.8)    & (0.5, 2.7)  & 1$^b$,31$^d$   \\
82 & NGC 4710    & 18.7 $\pm$ 2.5$^*$  & 1125\phantom{00000}    & 158\phantom{00000}    & 90 $\pm$ 27  & 7.8 $\pm$ 0.05  & (7.8, 8.1)    & (0.5, 2.7)    & 14$^b$,31$^d$   \\
83 & NGC 4736    & 21.3 $\pm$ 2.0$^*$  & 279\phantom{00000}    & 122\phantom{00000}    & 133 $\pm$ 19  & 6.8 $\pm$ 0.02  & 7.0 $\pm$ 0.04  & 0.7 $\pm$ 0.09  & 6$^{a,\dagger}$   \\
84 & MRK 231    & 8.1 $\pm$ 0.8\phantom{0}  & 12690 $\pm$ 10  & 215 $\pm$ 24  & 35 $\pm$\phantom{0} 9  & 9.5 $\pm$ 0.04  & 9.9 $\pm$ 0.06  & 0.4 $\pm$ 0.06  & 6,16 $^a$   \\
85 & NGC 4818    & 42.4 $\pm$ 1.1\phantom{0}  & 1055 $\pm$\phantom{0} 2  & 123 $\pm$\phantom{0} 4  & 324 $\pm$ 26  & 8.1 $\pm$ 0.01  & 8.1 & 1 & 8 $^a$   \\
86 & NGC 4826    & 86.8 $\pm$ 5.2$^*$  & 430\phantom{00000}    & 214\phantom{00000}    & 340 $\pm$ 61  & 7.8 $\pm$ 0.02  & (7.8, 8.1)  & (0.6, 2.0)  & 1$^b$,23$^d$  \\
87 & NGC 4941    & 6.7 $\pm$ 0.7\phantom{0}  & 1105 $\pm$\phantom{0} 3  & 56 $\pm$\phantom{0} 7  & 113 $\pm$ 19  & 7.3 $\pm$ 0.19  & $>$ 6.7    & $>$ 0.4  & 11 $^d$   \\
88 & NGC 5033    & 16.7 $\pm$ 1.1\phantom{0}  & 879 $\pm$\phantom{0} 8  & 240 $\pm$ 16  & 66 $\pm$ 13  & 7.5 $\pm$ 0.03  & 7.6 $\pm$ 0.03  & 0.8 $\pm$ 0.08  & 3 $^a$  \\
89 & NGC 5055    & 25.8 $\pm$ 3.5$^*$  & 595\phantom{00000}    & 190\phantom{00000}    & 99 $\pm$ 46  & 7.5 $\pm$ 0.10  & 8 & 0.4 & 8 $^a$   \\
90 & UGC 8335    & 3.6 $\pm$ 0.6$^*$  & 9371\phantom{00000}    & 154\phantom{00000}    & 15 $\pm$\phantom{0} 6  & 8.9 $\pm$ 0.05  & $<$ 9.6    & $<$ 1.2  & 10 $^c$  \\
91 & UGC 8335B    & 8.0 $\pm$ 1.2$^*$  & 9337\phantom{00000}    & 106\phantom{00000}    & 32 $\pm$ 14  & 9.2 $\pm$ 0.08  & $<$ 9.7    & $<$ 2.7  & 10 $^c$  \\
92 & Arp 193    & 20.5 $\pm$ 1.7$^*$  & 6967\phantom{00000}    & 240\phantom{00000}    & 64 $\pm$ 14  & 9.4 $\pm$ 0.04  & 9.6 $\pm$ 0.09  & 0.6 $\pm$ 0.12  & 9 $^a$  \\
93 & NGC 5194    & 44.4 $\pm$ 2.8\phantom{0}  & 422 $\pm$\phantom{0} 2  & 61 $\pm$\phantom{0} 5  & 681 $\pm$ 87  & 7.5 $\pm$ 0.03  & 7.5 $\pm$ 0.05  & 1.0 $\pm$ 0.13  & 6$^{a,\dagger}$  \\
94 & M 83        & 153.9 $\pm$ 3.1\phantom{0}  & 519 $\pm$\phantom{0} 1  & 112 $\pm$\phantom{0} 3  & 1290 $\pm$ 76  & 7.6 $\pm$ 0.01  & 7.6 & 1.1 $\pm$ 0.06  & 28 $^a$   \\
95 & NGC 5256    & 5.7 $\pm$ 0.9$^*$  & 8369\phantom{00000}    & 278\phantom{00000}    & 18 $\pm$\phantom{0} 5  & 9.0 $\pm$ 0.05  & $<$ 9.8    & $<$ 1.2  & 10 $^c$   \\
96 & NGC 5257    & 18.2 $\pm$ 1.5\phantom{0}  & 6787 $\pm$ 13  & 312 $\pm$ 29  & 55 $\pm$ 16  & 9.3 $\pm$ 0.04  & $>$ 9.3    & $>$ 0.5  & 13 $^d$  \\
97 & NGC 5258    & 27.0 $\pm$ 1.1\phantom{0}  & 6806 $\pm$\phantom{0} 8  & 347 $\pm$ 16  & 73 $\pm$ 12  & 9.5 $\pm$ 0.02  & $>$ 9.3    & $>$ 0.7  & 13 $^d$  \\
98 & NGC 5273    & $<$ 0.9\phantom{0}  & \nodata\phantom{0000} & ( 100 )    & ( 11 ) & $<$ 6.4   & $>$ 6.4   & \nodata & 11 $^d$  \\
99 & MRK 273    & 20.1 $\pm$ 0.7$^*$  & 11355\phantom{00000}    & 471\phantom{00000}    & 35 $\pm$\phantom{0} 6  & 9.8 $\pm$ 0.02  & 9.8 $\pm$ 0.02  & 0.9 $\pm$ 0.05  & 5 $^a$   \\
100 & NGC 5347    & 12.0 $\pm$ 1.4\phantom{0}  & 2382 $\pm$\phantom{0} 4  & 78 $\pm$ 10  & 144 $\pm$ 28  & 8.2 $\pm$ 0.08  & (7,7, 8.5)    & (1.7, 3.9)  & 7$^c$,11$^d$  \\
101 & NGC 5666    & 9.2 $\pm$ 1.8\phantom{0}  & 2321 $\pm$ 29  & 276 $\pm$ 51  & 31 $\pm$ 17  & 8.0 $\pm$ 0.08  & 8.2 $\pm$ 0.02  & 0.7 $\pm$ 0.15  & 20 $^a$   \\
102 & Arp 302S    & 8.5 $\pm$ 1.5$^*$  & 9774\phantom{00000}    & 232\phantom{00000}    & 27 $\pm$ 11  & 9.3 $\pm$ 0.07  & $<$ 9.6    & $<$ 3.4  & 10 $^c$  \\
    & Arp 302 center & $<$ 4.9\phantom{0}  & \nodata\phantom{0000} & ( 400 )    & ( 22 ) & $<$ 9.0 & \nodata  & \nodata & \nodata  \\
103 & Arp 302N    & 18.1 $\pm$ 1.1$^*$  & 10281\phantom{00000}    & 548\phantom{00000}    & 21 $\pm$\phantom{0} 9  & 9.6 $\pm$ 0.07  & $<$ 10.4    & $<$ 1.3  & 10 $^c$   \\
104 & NGC 5866    & $<$ 8.6\phantom{0}  & \nodata\phantom{0000} & ( 200 )    & ( 78 ) & $<$ 7.0   & 7.2 & $<$   0.6  & 12 $^a$   \\
105 & NGC 5907   & 8.4 $\pm$ 0.7\phantom{0}  & 663 $\pm$\phantom{0} 3  & 82 $\pm$\phantom{0} 9  & 96 $\pm$ 19  & 7.0 $\pm$ 0.04  & 7.5 $\pm$ 0.02  & 0.3 $\pm$ 0.03  & 3 $^a$   \\
106 & Mrk 848    & 4.2 $\pm$ 0.4\phantom{0}  & 12111 $\pm$\phantom{0} 4  & 88 $\pm$ 11  & 44 $\pm$\phantom{0} 7  & 9.1 $\pm$ 0.05  &  9.4    &  0.5  & 19 $^a$  \\
107 & NGC 5953    & 21.5 $\pm$ 2.2$^*$  & 2003\phantom{00000}    & 147\phantom{00000}    & 112 $\pm$ 40  & 8.3 $\pm$ 0.04  & (8.4, 8.6)  & (0.5, 2.3)  & 13 $^d$  \\
108 & Arp 220    & 58.5 $\pm$ 1.9$^*$  & 5427\phantom{00000}    & 379\phantom{00000}    & 129 $\pm$\phantom{0} 8  & 9.6 $\pm$ 0.02  & 9.9 $\pm$ 0.04  & 0.5 $\pm$ 0.05  & 7,15 $^a$  \\
109 & NGC 6240    & 74.9 $\pm$ 2.2$^*$  & 7394\phantom{00000}    & 355\phantom{00000}    & 186 $\pm$ 21  & 10.0 $\pm$ 0.01  & 9.9 $\pm$ 0.09  & 1.1 $\pm$ 0.22  & 15,4 $^a$   \\
110 & 17208-0014    & 24.6 $\pm$ 1.4\phantom{0}  & 12857 $\pm$ 11  & 386 $\pm$ 24  & 60 $\pm$\phantom{0} 9  & 10.0 $\pm$ 0.03  & 9.9 $\pm$ 0.09  & 1.2 $\pm$ 0.26  & 15 $^a$  \\
111 & Arp 293    & 13.3 $\pm$ 1.8$^*$  & 5431\phantom{00000}    & 282\phantom{00000}    & 56 $\pm$ 14  & 9.0 $\pm$ 0.05  & $<$ 9.5    & $<$ 1.4  & 1 $^b$  \\
112 & NGC 6524   & 6.6 $\pm$ 1.1\phantom{0}  & 5659 $\pm$ 10  & 125 $\pm$ 24  & 49 $\pm$ 14  & 8.7 $\pm$ 0.07  & \nodata & \nodata & \nodata  \\
113 & NGC 6670B    & 19.8 $\pm$ 1.8$^*$  & 8618\phantom{00000}    & 360\phantom{00000}    & 58 $\pm$ 20  & 9.5 $\pm$ 0.07  & $<$ 9.9    & $<$ 2.8  & 10 $^c$   \\
  & NGC 6670    & 8.3 $\pm$ 0.7$^*$  & 8688\phantom{00000}    & 268\phantom{00000}    & 45 $\pm$ 19  & 9.2 $\pm$ 0.04  & $<$ 10.2    & $<$ 0.7  & 10 $^c$   \\
114 & NGC 6670A    & 16.2 $\pm$ 1.5$^*$  & 8702\phantom{00000}    & 307\phantom{00000}    & 27 $\pm$\phantom{0} 8  & 9.5 $\pm$ 0.05  & $<$ 9.8    & $<$ 3.5 $^a$  & 10 $^c$   \\
115 & NGC 6814    & 2.1 $\pm$ 0.2\phantom{0}  & 1557 $\pm$\phantom{0} 7  & 119 $\pm$ 13  & 17 $\pm$\phantom{0} 4  & 7.1 $\pm$ 0.06  & (7.3, 8.1)  & (0.3, 0.5)  & 1$^b$,11$^d$   \\
116 & NGC 6946    & 132.3 $\pm$ 1.8\phantom{0}  & 60 $\pm$\phantom{0} 1  & 150 $\pm$\phantom{0} 2  & 827 $\pm$ 56  & 7.7 $\pm$ 0.01  & 7.8 $\pm$ 0.03  & 0.9 $\pm$ 0.07  & 6 $^a$   \\
117 & NGC 7013    & 10.5 $\pm$ 0.8\phantom{0}  & 829 $\pm$ 13  & 352 $\pm$ 30  & 28 $\pm$\phantom{0} 7  & 7.2 $\pm$ 0.03  & 7 & 1.4 & 12 $^a$   \\
118 & NGC 7077    & 0.8 $\pm$ 0.2\phantom{0}  & 1015 $\pm$ 16  & 222 $\pm$ 31  & 19 $\pm$\phantom{0} 7  & 6.4 $\pm$ 0.06  & 6.3 & 1.1 $\pm$ 0.38  & 21 $^a$   \\
119 & NGC 7217    & 8.1 $\pm$ 0.7\phantom{0}  & 924 $\pm$ 12  & 269 $\pm$ 25  & 28 $\pm$\phantom{0} 8  & 7.2 $\pm$ 0.04  & 7.3 $\pm$ 0.02  & 0.8 $\pm$ 0.08  & 3 $^a$   \\
120 & NGC 7331    & 7.2 $\pm$ 0.6$^*$  & 884\phantom{00000}    & 50\phantom{00000}    & 74 $\pm$ 15  & 7.3 $\pm$ 0.03  & 8.0 $\pm$ 0.01  & 0.2 $\pm$ 0.02  & 3 $^a$   \\
  & NGC 7331A    & 12.6 $\pm$ 1.0$^*$  & 903\phantom{00000}    & 145\phantom{00000}    & 86 $\pm$ 25  & 7.6 $\pm$ 0.05  & 8.0 $\pm$ 0.01  & 0.3 $\pm$ 0.03  & 3 $^a$  \\
121 & NGC 7465    & 6.3 $\pm$ 0.9\phantom{0}  & 1970 $\pm$ 10  & 127 $\pm$ 17  & 47 $\pm$ 16  & 7.8 $\pm$ 0.06  & $<$ 8.2    & $<$ 1.6  & 1 $^b$  \\
122 & NGC 7469    & 35.2 $\pm$ 1.3\phantom{0}  & 4949 $\pm$\phantom{0} 5  & 249 $\pm$ 10  & 133 $\pm$ 22  & 9.3 $\pm$ 0.02  & (8.8, 9.4)  & (1.5, 3.4)  & 1$^b$,11$^d$   \\
123 & NGC 7479    & 20.1 $\pm$ 1.3\phantom{0}  & 2381 $\pm$\phantom{0} 8  & 238 $\pm$ 19  & 79 $\pm$ 15  & 8.4 $\pm$ 0.03  & $<$ 9.1    & $<$ 1.6  & 1 $^b$   \\
124 & NGC 7541    & 17.8 $\pm$ 0.9\phantom{0}  & 2673 $\pm$\phantom{0} 4  & 144 $\pm$\phantom{0} 9  & 116 $\pm$ 20  & 8.3 $\pm$ 0.02  & 8.5 & 0.6 & 8 $^{a,\dagger}$   \\
125 & NGC 7679    & 17.2 $\pm$ 1.4$^*$  & 5110\phantom{00000}    & 285\phantom{00000}    & 55 $\pm$ 29  & 9.0 $\pm$ 0.05  & $<$ 9.0    & (1.1, 2.4)  & 12 $^e$  \\
\label{tbl:para}
\end{longtable}
{
The columns contain the following information:
Col.(1): Sequence number.
Col.(2):  Galaxy name.
Cols.(3)--(6):  The CO(3--2) line
intensity ($I_{32}$ = $\int T_{\rm mb}\,{\rm d}v$), LSR velocity, 
width (FWHM), and main beam brightness temperature, respectively, 
with corresponding standard errors from Gaussian fits or from 
$^*$moments in the case of non-Gaussian line shapes. Non-detections
are listed with their rms noise level $\sigma$ (Col.(6) values in
brackets) at a channel spacing of $\delta v$~$\sim$~6.6~\kms. In 
these cases the full width of the CO(1--0) line ($\Delta$$V$$_{\rm 10}$, 
Col.(5) values in brackets) was estimated from spectra taken from the 
literature or was set to 400~\kms; corresponding upper limits 
for $I_{32}$ were obtained using $I_{32}$~$<$~
$3\sigma\times(\Delta\,V_{\rm 10}\times\delta\,v)^{1/2}$ (Col.(3)).
Col.(7): The CO(3--2) luminosity calculated with Eq.(1)
(see \S\,4.1) for a beam of $\Theta_{\rm mb}$ = 22$''$.
Col.(8): The CO(1--0) luminosity calculated with an equation
similar to that of $L_{\rm CO(3-2)}$ for galaxies with IRAM-30m
CO(1--0) data available in the literature (listed in Col(10));
otherwise either upper or lower limits are given, depending on the
telescope used for CO(1--0).
Col.(9): The CO(3--2)/CO(1--0) line intensity ratio $R_{31}$ and
its standard deviation for galaxies with IRAM-30m CO(1--0) data
available in the literature (listed in Col(10)); otherwise either
upper or lower limits are given, depending on the telescope used
for the CO(1--0) data.
Col.(10): References for the CO(1--0) data:
1. \citet{you95}; 2. \citet{alb04}; 3.\citet{bra93}; 4.
\citet{com91}; 5. \citet{gs04b}; 6. \citet{ger00}; 7.
\citet{hec89}; 8. \citet{mau99}; 9. \citet{rad91}; 10.
\citet{san91}; 11. \citet{vil98}; 12. \citet{wik89}; 13.
\citet{yao03}; 14. \citet{ken88}; 15. \citet{sol97}; 16.
\citet{reu96}; 17. \citet{sch07}; 18. \citet{cas89}; 19.
\citet{chi92}; 20. \citet{wik95}; 21. \citet{sag92}; 22.
\citet{reu93}; 23. \citet{nis01}; 24. \citet{gol93}; 25.
\citet{han90}; 26. \citet{kun07}; 27. \citet{com07}; 28.
\citet{baa08}; 29. \citet{sol92}; 30. \citet{gre96}; 31.
\citet{kom08}. Telescopes used are $^a$ IRAM-30m; $^b$ FCRAO-14m;
$^c$ NRAO--12m; $^d$ NRO-45m; $^e$ OSO-20m. $^\dagger$: The
CO(1--0) data was taken at an offset position $>$ 5$''$ relative
to our CO(3--2) position. }
}

\clearpage
\begin{deluxetable}{lcccccc}
\tabletypesize{\scriptsize}
\tablecaption{$^{12}$CO(3--2)/(1--0) integrated line intensity ratios.
\label{tbl:ratios}}
\tablehead{
\multicolumn{1}{l}{galaxy} &
\multicolumn{6}{c}{$R_{\rm 31}$ statistics} \\
\cline{2-7}
\multicolumn{1}{l}{type} & \multicolumn{1}{c}{N} &
\multicolumn{1}{c}{mean\tablenotemark{d}}  & $\sigma$\tablenotemark{e}   & \multicolumn{1}{c}{median} &
\multicolumn{1}{c}{min} & \multicolumn{1}{c}{max} } \startdata
(activity) & \multicolumn{5}{c}{} \\
\hline
normal    &  7 & 0.61 $\pm$ 0.16 & 0.42 & 0.44 & 0.26 & 1.51 \\
LINER\tablenotemark{a}     &  20 & 0.65 $\pm$ 0.08 & 0.36 & 0.57 & 0.19 & 1.57 \\
Seyfert\tablenotemark{a}   &  12 & 0.82 $\pm$ 0.12 & 0.43 & 0.75 & 0.27 & 1.79 \\
`pure' AGN\tablenotemark{b}  &  18 & 0.78 $\pm$ 0.09 & 0.37 & 0.70 & 0.19 & 1.57 \\
starburst\tablenotemark{c} &  25 & 0.89 $\pm$ 0.11 & 0.53 & 0.71 & 0.25 & 1.93 \\
(U)LIRG   &  10 & 0.96 $\pm$ 0.14 & 0.45 & 0.99 & 0.37 & 1.63 \\
\hline
(bar) & \multicolumn{5}{c}{} \\
\hline
SA        &  18 & 0.62 $\pm$ 0.10 & 0.41 & 0.50 & 0.19 & 1.71 \\
SAB       &  16 & 0.80 $\pm$ 0.11 & 0.44 & 0.72 & 0.25 & 1.93 \\
SB        &  12 & 0.88 $\pm$ 0.15 & 0.53 & 0.67 & 0.30 & 1.63 \\
\hline
\\
 Total     &  61 & 0.81 $\pm$ 0.06 & 0.46 & 0.69 & 0.19 & 1.93 \\
\enddata
\tablenotetext{a}{Including starbursts but not (U)LIRG overlaps.}
\tablenotetext{b}{See \S.~\ref{section:classification} for
details.} \tablenotetext{c}{Excluding (U)LIRGs.}
\tablenotetext{d}{The mean value and its standard error.}
\tablenotetext{e}{Standard deviation of an individual target.}
\end{deluxetable}

\clearpage
\begin{figure*}
\centering
\includegraphics[scale=0.7,angle=0]{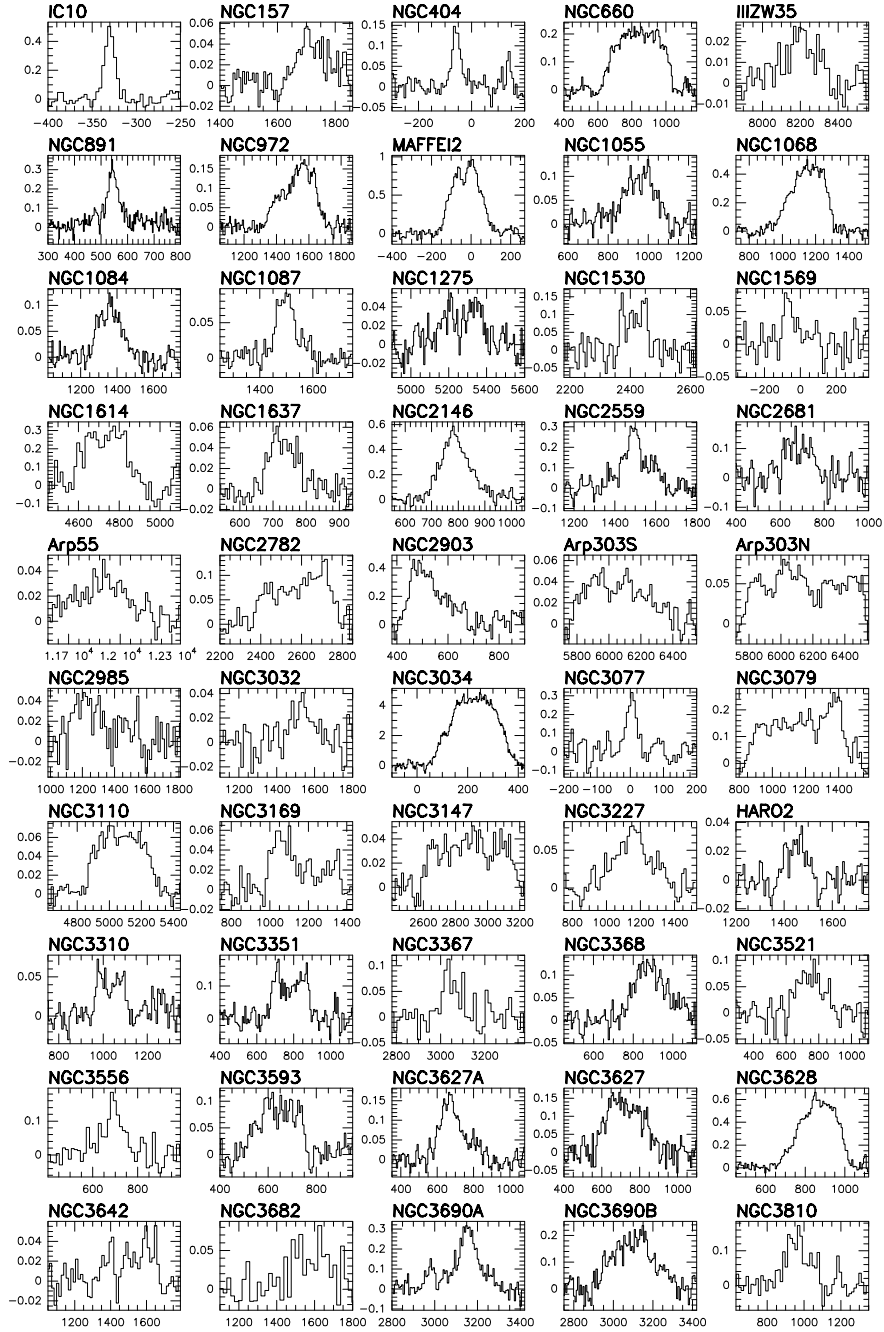}
\caption[]{CO $J$ = 3--2 spectra. The velocity scale corresponds
to Local Standard of Rest in units of \kms. The intensity is
displayed in units of main beam brightness temperature (K).}
\label{fig:co-spectra}
\end{figure*}

\addtocounter{figure}{-1}
\begin{figure*}
\centering
\includegraphics[scale=0.7,angle=0]{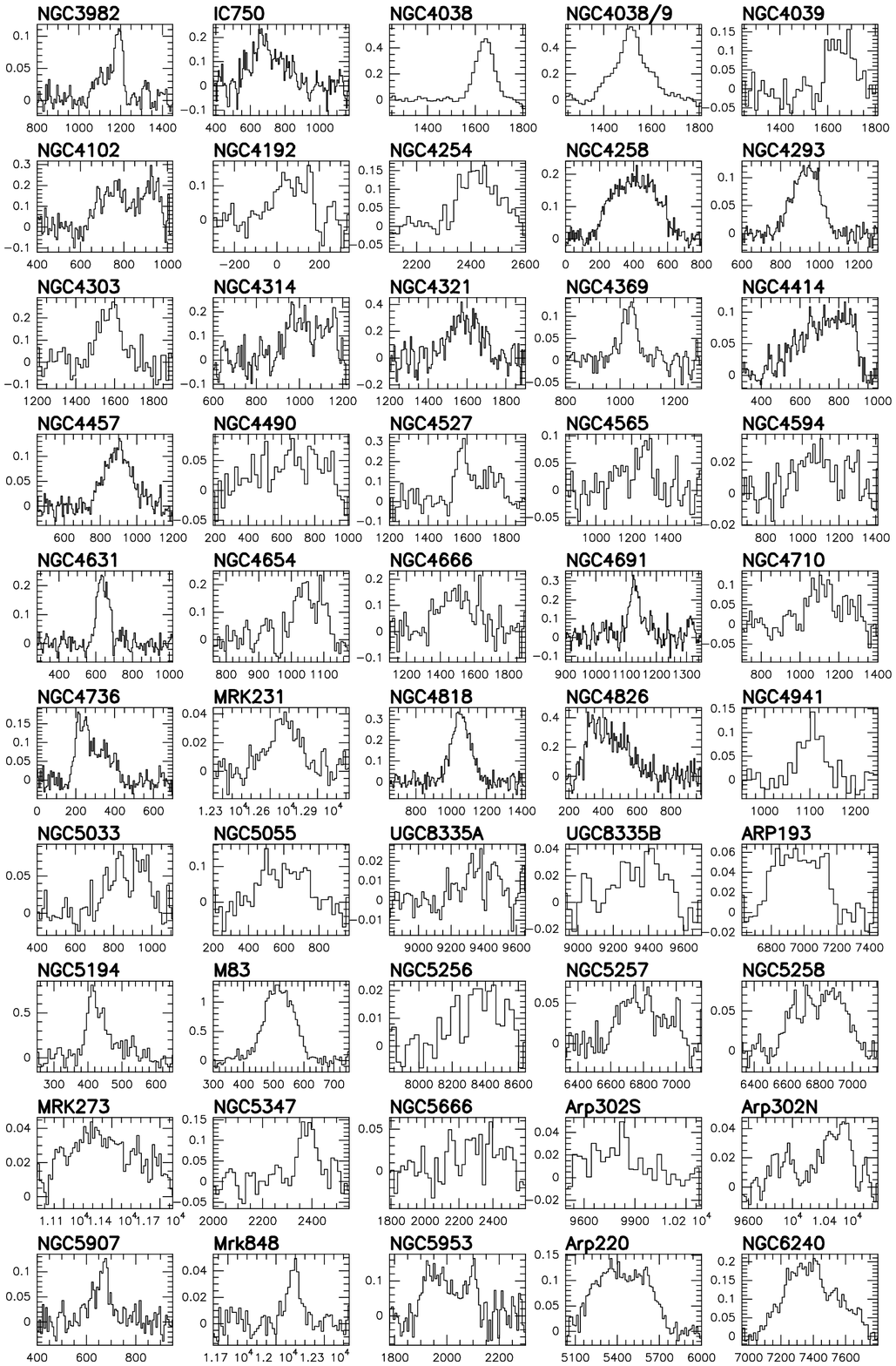}
\caption{({\it Continued})}
\end{figure*}

\addtocounter{figure}{-1}
\begin{figure*}
\centering
\includegraphics[scale=0.7,angle=0]{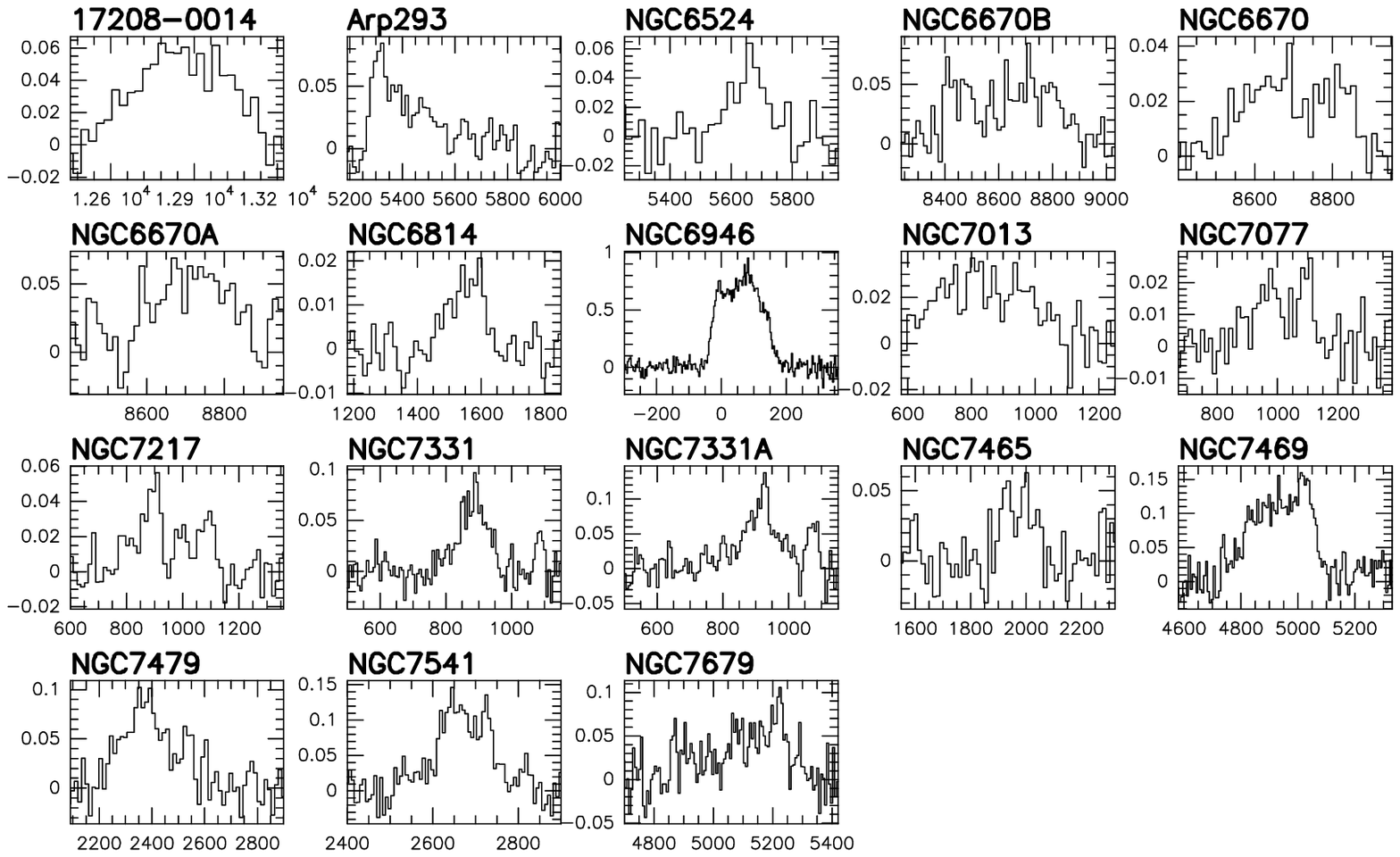}
\caption{({\it Continued})}
\ \ \ \ \ \\
\ \ \ \ \ \\
\end{figure*}

\begin{figure}
\centering
\includegraphics[angle=0,width=8cm]{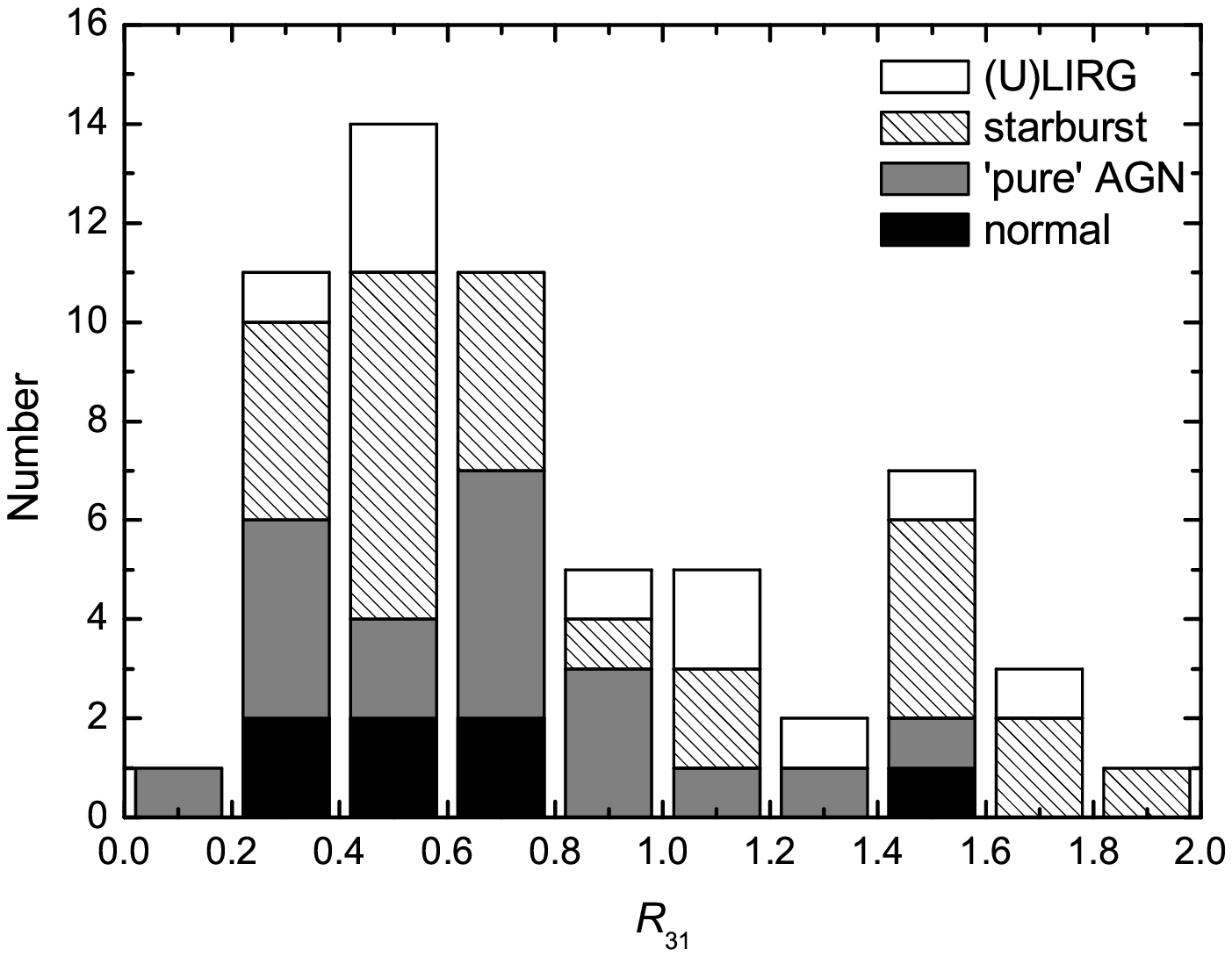}
\caption[]{Number distribution of the matching beam line intensity
ratio $R_{\rm 31}$ for 61 galaxies from our sample (see \S\,5.1.2
for details).} \label{fig:N-R31}
\end{figure}

\clearpage
\begin{figure*}
\centering
\includegraphics[angle=0,width=16.5cm]{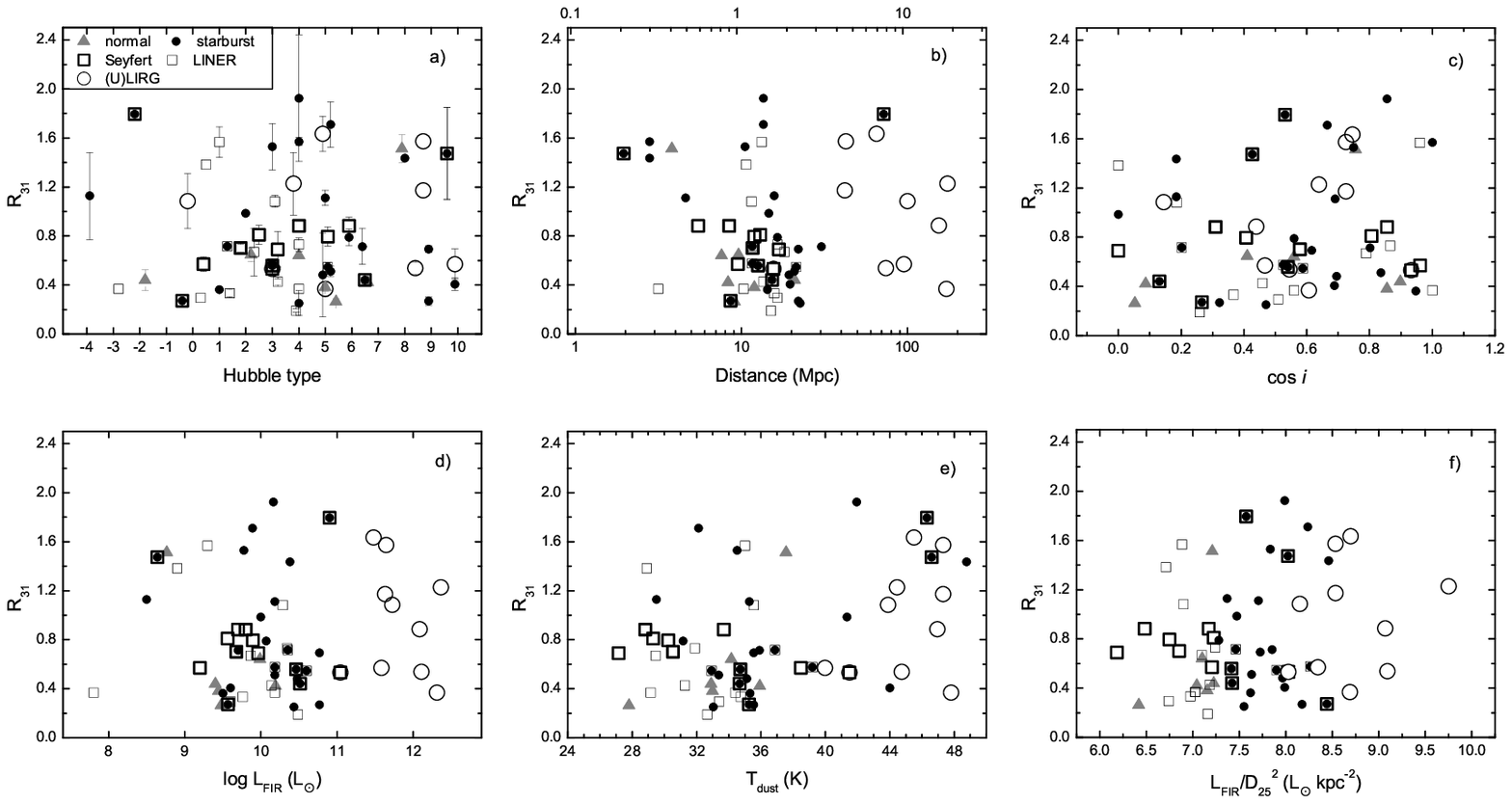}
\caption[]{$R_{\rm 31}$ versus a) Hubble type, b) distance (the
lower X-axis) or projected beam size (the upper X-axis in kpc) c)
cosine of the inclination, d) FIR luminosity, e) dust temperature,
and f) FIR luminosity per unit area ($L_{\rm FIR}$/$D_{\rm 25}^2$)
of sample galaxies with galaxy types being indicated in a). The
error bars are removed for clarity in the panels b) -- f).}
\label{fig:R31-all}
\end{figure*}

\clearpage
\begin{figure}
\centering
\includegraphics[width=12.8cm]{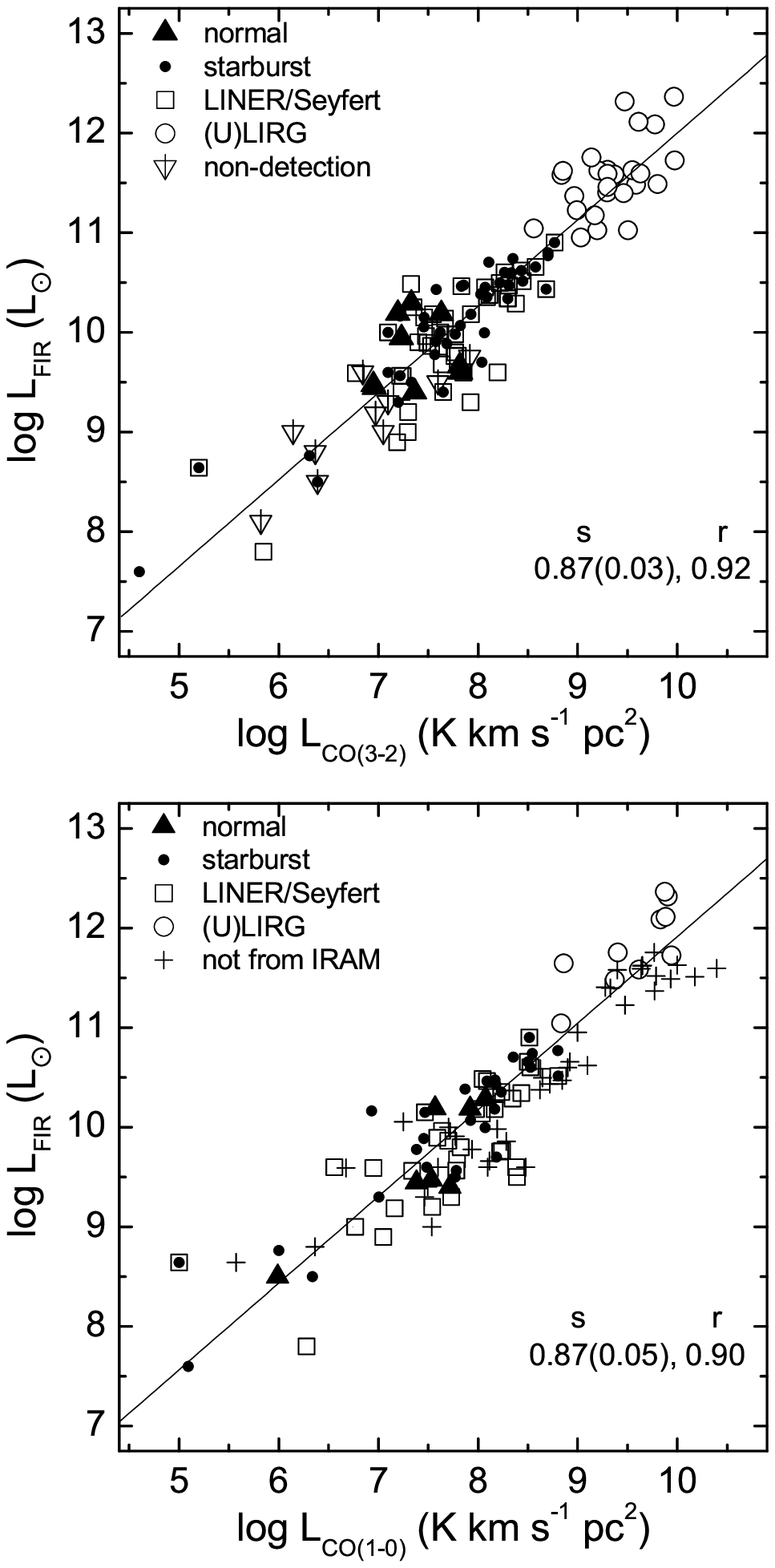}
\caption[]{Log-log plot of the correlation between nuclear CO and
global FIR luminosity for CO(3--2) (upper panel) and CO(1--0)
(lower panel). Both CO line luminosities cover the central 22$''$
region, except for the crosses representing CO(1--0) data with the
larger ($\sim$50$''$) beams of the FCRAO-14m or NRAO-12m telescopes.
Straight lines show linear regression fits to the unweighted data.
Slopes (s) and correlation coefficients (r) are given at the lower
right corner of each panel. Only our CO(3--2) detections and
CO(1--0) data from the IRAM-30\,m are included in the fits.
Non-detections (upper limits) and data from the FCRAO and NRAO
were not considered. IC\,10, the isolated dot at the lower left
corner of each panel, is also not part of the fits (see
\S\,5.2.2).} \label{fig:l-l-plot}
\end{figure}

\clearpage
\begin{figure}
\centering
\includegraphics[width=20.0cm]{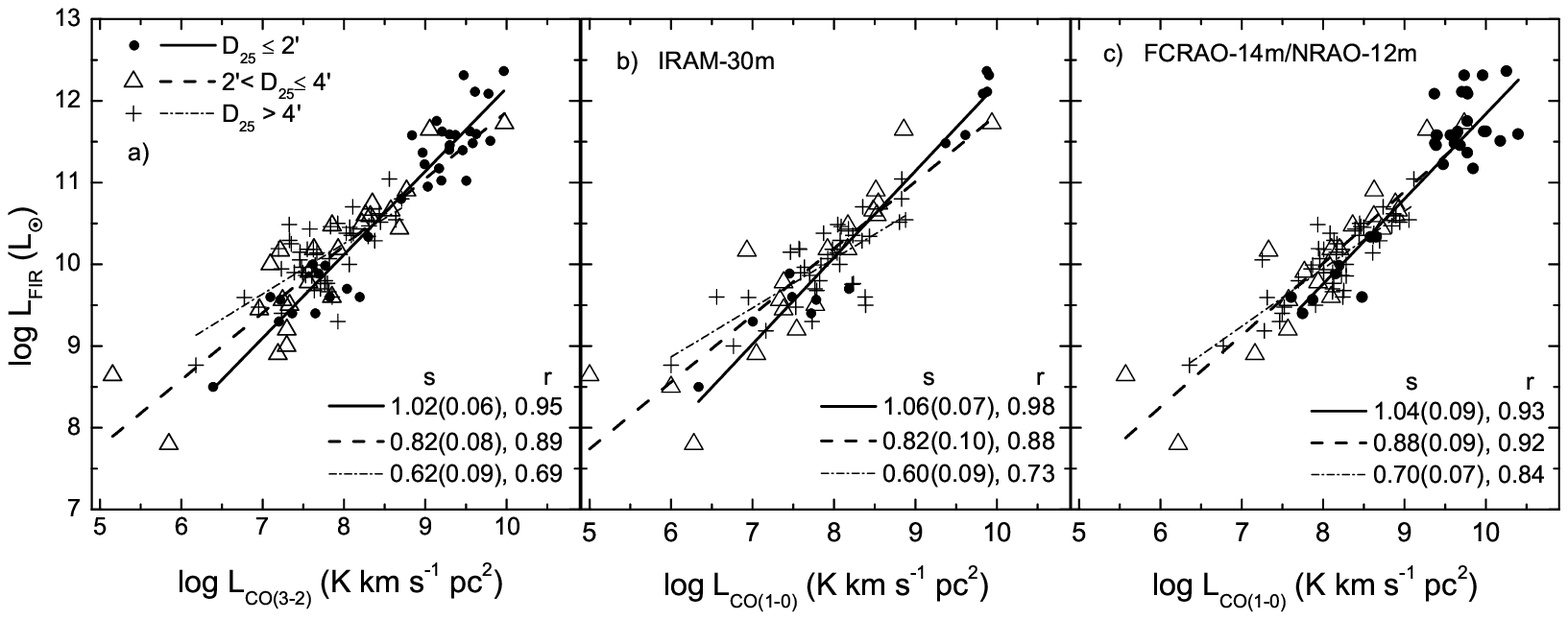}
\vspace{-18.0cm}
\caption[]{Log-log correlation of a) $L_{\rm CO(3-2)}$ and $L_{\rm
FIR}$ (this work), b) $L_{\rm CO(1-0)}$ and $L_{\rm FIR}$ (CO from
the IRAM-30m telescope), and c) $L_{\rm CO(1-0)}$ and $L_{\rm
FIR}$ (CO from the FCRAO-14m and NRAO-12m antennas), with our
sample galaxies being divided into three groups characterized by
their optical angular sizes ($D_{\rm 25}$): 1) $D_{\rm
25}$~$\leq$~2$'$ (filled circles, solid lines denoting the
corresponding linear regression fit), 2)
2$'$~$<$~$D_{\rm 25}$~$\leq$~4$'$ (empty triangles, dashed lines),
and 3) 4$'$~$<$~$D_{\rm 25}$~$\leq$~18$'$ (crosses, dash-dotted
lines). The corresponding slopes ($s$) and correlation coefficients
($r$) are also given.} \label{fig:l-l-d25}
\end{figure}

\clearpage
\begin{figure}
\centering
\includegraphics[angle=0,width=8cm]{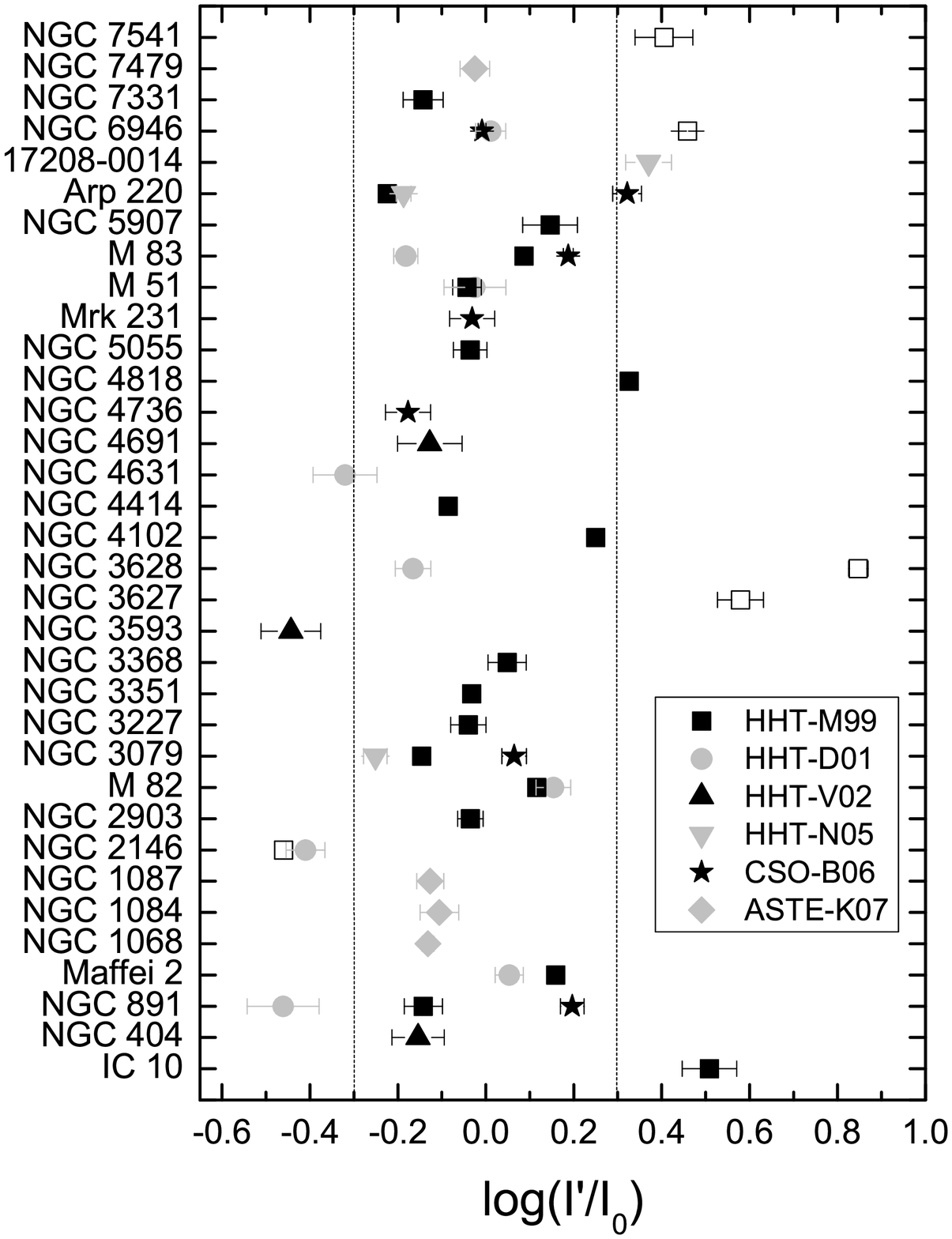}
\caption[]{A comparison of our integrated CO(3--2) intensities
with previously published results. The horizontal
axis gives the logarithmic intensity deviation, where $I'$ denotes
the integrated CO(3--2) intensity taken from the literature,
while $I_0$ is from this work. Data points measured at nominal
position offsets $\ga$~5$''$ are represented as open squares.
A pair of vertical lines marks
the $\pm$0.3\,dex deviation limits. For the references given in
the lower right box, see \S A.1.} \label{fig:consistency}
\end{figure}

\clearpage
\begin{figure}
\centering
\includegraphics[angle=0,width=8cm]{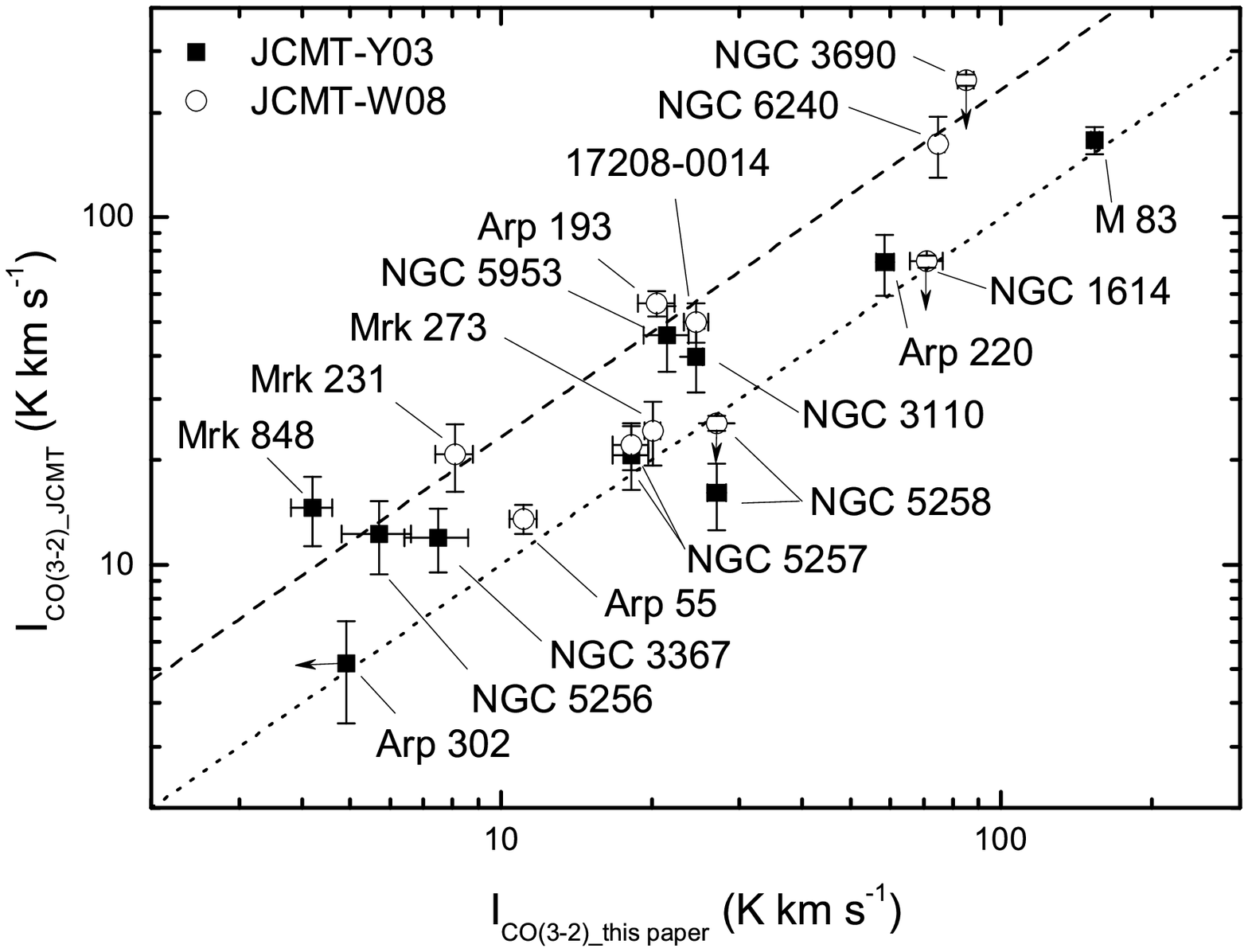}
\caption[]{A comparison of CO(3--2) integrated intensities for
galaxies observed with the JCMT--15m (by \citet[JCMT-Y03]{yao03}
and \citet[JCMT-W08]{wil08}) and with the HHT--10m in this paper.
Two straight lines denote the theoretical relationship of
intensities obtained with the two telescopes assuming point-like
(dashed) and uniformly-extended (dotted) structures with respect
to the observing beams. } \label{fig:JCMT}
\end{figure}
\label{lastpage}

\end{document}